\documentclass[twocolumn]{autart}    % Enable this line and disable the 
\usepackage{graphics}
\usepackage{times}
\usepackage{amsmath,amssymb,amsfonts}
\allowdisplaybreaks
\usepackage{algorithm}
\usepackage{xcolor}
\usepackage{siunitx}
\usepackage{bm}
\usepackage{tikz}
\usepackage{pgfplots}
\usepackage{hyperref}
\usepackage{bookmark}
\usepackage{booktabs}

\pgfplotsset{compat=1.18}
\usetikzlibrary{arrows,shapes,backgrounds,patterns,fadings,decorations,
	positioning,external,arrows.meta,pgfplots.fillbetween, pgfplots.groupplots}

\definecolor{oiOrange}{HTML}{E69F00}
\definecolor{oiSkyBlue}{HTML}{56B4E9}
\definecolor{oiGreen}{HTML}{009E73}
\definecolor{oiYellow}{HTML}{F0E442}
\definecolor{oiBlue}{HTML}{0072B2}
\definecolor{oiVermilion}{HTML}{D55E00}
\definecolor{oiPurple}{HTML}{CC79A7}
\definecolor{oiBlack}{HTML}{000000}

\let\classAND\AND
\let\AND\relax
\usepackage{algorithmic}

\let\AND\classAND
\AtBeginEnvironment{algorithmic}{\let\AND\algoAND}

\begin{document}

\begin{frontmatter}

% Running title for regular 
%\runtitle{Bayesian Optimal Control from Input--Output Data}

\title{A Bayesian Framework for Safe Optimal Control of Unknown Systems from Input--Output Data\thanksref{footnoteinfo}}

\thanks[footnoteinfo]{The material in this article was partially presented at the 22nd European Control Conference, June~25--28, 2024, Stockholm, Sweden. Corresponding author Robert Lefringhausen. 
	%Tel. +49 (89) 289-25723. 
	%Fax +49 (89) 289-25724.
 }

\author{Robert Lefringhausen}\ead{robert.lefringhausen@tum.de},
\author{Annalena Daniels\thanksref{NTFE}}\ead{annalena.daniels@tum.de}, 
\author{Michael Fink\thanksref{NTFE}}\ead{michael.fink@tum.de},
\author{Sandra Hirche}\ead{hirche@tum.de}  
\address{School of Computation, Information and Technology, Technical University of Munich, 
	% Arcisstrasse 21, 80333 Munich, 
	Germany}
	
\thanks[NTFE]{Present address: NTFE GmbH, Munich, Germany.}

% Five to ten keywords, 
% chosen from the IFAC 
% keyword list or with the 
% help of the Automatica 
% keyword wizard 

% Abstract of not more than 200 words.
\begin{abstract}
As control engineering is applied to increasingly complex systems, learning-based control, which uses empirical data to reduce the reliance on first-principles models, has emerged as a promising paradigm. However, such approaches must cope with nonlinear and stochastic dynamics, the absence of state measurements, and scarce data, while ensuring safe operation under constraints, which requires rigorous uncertainty quantification. Whereas existing methods treat these challenges in isolation, this article proposes a Bayesian framework that jointly addresses them. A prior in state-space form is updated from input--output data using particle Markov chain Monte Carlo methods, yielding a representation of joint uncertainty over dynamics and latent states. Building on this representation, a validation procedure is developed that provides probabilistic guarantees for constraint satisfaction under arbitrary causal control laws. For control synthesis, a formulation penalizing the conditional value-at-risk of the constraint function is proposed, together with an adaptive calibration mechanism that adjusts the performance--robustness trade-off to achieve a desired robustness level. The probability of successful certification is characterized theoretically. The framework is extended to a receding-horizon setting, where posterior updates from new measurements allow the controller to progressively reduce uncertainty and improve performance. The approach is demonstrated on the optimal control of a vertical farm.
\end{abstract}

\end{frontmatter}

\section{Introduction} 
\label{sec:introduction}

Learning-based optimal control aims to enable the high-performance operation of systems with unknown dynamics by learning models from empirical data. This paradigm is particularly relevant in applications where pure first-principles modeling is infeasible or prohibitively expensive, such as soft robotic actuators, systems involving human interaction, and biological processes. In real-world applications, several challenges arise jointly that fundamentally shape the learning-based control problem:
\begin{enumerate}
	\item \textbf{Nonlinear and stochastic dynamics:} Real-world systems are inherently nonlinear and affected by both process noise, which alters the system trajectory, and measurement noise, which only corrupts observations.
	\item \textbf{Absence of state measurements:} In many applications, the system state is latent, i.e., it cannot be measured directly due to limited sensing capabilities. Consequently, methods must operate solely on input--output data.
	\item \textbf{Scarce data:} In engineering applications, data are often scarce due to costly and time-consuming experiments, making the incorporation of structural knowledge from first principles essential for data efficiency and generalization.
	\item \textbf{Rigorous safety guarantees:} In safety-critical applications, constraint satisfaction must be ensured with high confidence despite the challenges posed by~\mbox{(1)--(3)}. This necessitates the rigorous quantification of both the irreducible aleatoric uncertainty introduced by process and measurement noise and the reducible epistemic uncertainty about the dynamics and latent states that arises from limited data.
\end{enumerate}
Despite the practical importance of these challenges, existing approaches address individual aspects only, while their combined treatment remains largely unresolved. In the following, we briefly review existing approaches and their limitations with respect to these challenges.

\newpage
Data-driven control methods rooted in behavioral systems theory~\cite{berberich2025} formulate optimal control problems directly from input--output data without requiring measurements of the underlying system state. These approaches have primarily been developed for linear time-invariant systems, and although extensions to stochastic and nonlinear settings have been proposed, their guarantees rely on specific system classes or local approximations.

Nonlinear input--output models, such as nonlinear autoregressive exogenous (NARX) models, learn a mapping from past inputs and outputs to future system behavior. Bayesian formulations, for instance Gaussian process (GP) NARX models~\cite{maiworm2021}, systematically quantify both epistemic and aleatoric uncertainty through a posterior distribution over this mapping, potentially enabling control with probabilistic safety guarantees. However, input--output models lack an internal state and therefore cannot distinguish process noise from measurement noise, treating all stochastic variability as a single lumped disturbance. This can lead to biased and increasingly inaccurate multi-step predictions over the horizons relevant for optimal control~\cite{schoukens2019}. Moreover, prior knowledge from first principles, such as conservation laws or kinematic relations, is naturally expressed in terms of state variables and cannot be straightforwardly encoded in input--output mappings.

State-space models are a compelling foundation for learning-based optimal control, as they can describe nonlinear dynamics, inherently separate process and measurement noise, and allow the direct integration of prior knowledge from first principles. In their Bayesian formulation, for instance as GP state-space models, they further provide a principled framework for quantifying both epistemic and aleatoric uncertainty, which several works have exploited for robust learning-based control~\cite{scampicchio2025}. These approaches, however, rely on full state measurements, which are unavailable in many applications. Without them, learning the dynamics no longer reduces to fitting a function to observed state transitions, and predictions over a planning horizon depend on a current state that is itself unknown. The unknown dynamics and the latent states must therefore be estimated jointly. While several works provide point estimates of both~\cite{schoen2011,schiller2026}, quantifying the remaining uncertainty is substantially more challenging because the uncertainties of the dynamics and the latent states are coupled and their joint posterior is generally intractable. To the best of our knowledge, no existing control method for nonlinear systems accounts for this joint uncertainty and translates it into certifiable probabilistic guarantees, except for our preliminary work~\cite{lefringhausen2024}.

This article closes this gap by proposing a framework for the optimal control of nonlinear systems with unknown dynamics from input--output data that addresses all four challenges. To retain the advantages of Bayesian state-space models without requiring full state measurements, we use particle Markov chain Monte Carlo (PMCMC) methods~\cite{andrieu2010} to sample from the intractable joint posterior over dynamics and latent states. Together with the noise model, these samples provide a representation of both epistemic and aleatoric uncertainty. While PMCMC-based inference for nonlinear state-space models has been studied in the context of system identification~\cite{frigola2013,svensson2017,wigren2022}, we exploit this representation for control, with the following main contributions:
\begin{enumerate}
	\item[(i)] We develop a validation procedure that provides non-asymptotic probabilistic guarantees for constraint satisfaction under arbitrary causal control laws with respect to the posterior predictive distribution over dynamics, latent states, and future disturbances.
		
	\item[(ii)] We propose a synthesis--validation procedure that computes a certified input sequence by combining the validation procedure in~(i) with risk-sensitive optimal control using a conditional value-at-risk (CVaR) penalty. The weight of this penalty trades off performance and robustness, and an adaptive calibration mechanism adjusts it until the desired robustness level is certified. We further analyze its theoretical properties, linking the risk weight to the risk of constraint violation and quantifying the probability of successful certification.

	\item[(iii)] We extend the synthesis--validation procedure in~(ii) to a receding-horizon setting in which the posterior is updated between control cycles as new measurements become available, enabling the controller to progressively reduce uncertainty and improve performance. By exploiting the gradual evolution of the posterior, each cycle can be performed more efficiently than repeating inference and synthesis from scratch, while preserving the per-cycle probabilistic guarantees.
\end{enumerate}

This article extends our preliminary work~\cite{lefringhausen2024}, which proposes PMCMC-based open-loop optimal control with robustness guarantees from non-convex scenario theory~\cite{campi2021}. The present work replaces this conservative and computationally demanding formulation with a synthesis--validation framework in which the robustness level is an explicit design choice, extends it to a receding-horizon setting, and demonstrates it in a comprehensive case study.

\paragraph*{Outline}
After the problem formulation in Section~\ref{sec:problem_formulation} and a brief review of PMCMC methods in Section~\ref{sec:pmcmc}, Section~\ref{sec:validation} develops the validation procedure and its probabilistic guarantees. Section~\ref{sec:optimal_control} presents the CVaR-penalized control synthesis with adaptive calibration and its analysis, and Section~\ref{sec:receding_horizon} extends the framework to a receding-horizon setting. Section~\ref{sec:example} demonstrates the approach on the optimal control of a vertical farm, and Section~\ref{sec:conclusion} concludes the article.

\paragraph*{Notation}
Lowercase/uppercase bold symbols denote vectors/matrices, respectively. $\mathbb{R}$ denotes the set of real numbers, and $\mathbb{N}^0$ and $\mathbb{N}$ denote the sets of natural numbers with and without zero, respectively. For indices~$b \leq c$, $\bm{a}_{b:c}$ is the shorthand notation for $(\bm{a}_{b}, \dots, \bm{a}_{c})$. $\mathbb{P}(\cdot)$ denotes the probability of an event, $\mathbb{E}[\cdot]$ the expected value, and $p(\cdot)$ the probability density function.

\section{Problem Formulation}
\label{sec:problem_formulation}

Consider a nonlinear discrete-time system
\begin{subequations} \label{eq:system}
	\begin{align}
		\bm{x}_{t+1} &= \bm{f}_{\bm{\theta}}\left(\bm{x}_t,\bm{u}_t,\bm{v}_t \right)\\
		\bm{y}_{t} &= \bm{g}_{\bm{\theta}} \left(\bm{x}_t,\bm{u}_t \right)+\bm{w}_t,
	\end{align}
\end{subequations}
where $\bm{x}_t \in \mathbb{R}^{n_x}$ denotes the state, $\bm{u}_t \in \mathbb{R}^{n_u}$ the input, and $\bm{y}_t \in \mathbb{R}^{n_y}$ the output at time~$t\in\mathbb{N}^0$. The state~$\bm{x}_t$ is not directly observed, and only the output~$\bm{y}_t$ is available for feedback and learning. The vector~$\bm{\theta} \in \mathbb{R}^{n_\theta}$ collects all unknown model parameters. The state-transition function~$\bm{f}_{\bm{\theta}}(\cdot)$ and the observation function~$\bm{g}_{\bm{\theta}}(\cdot)$ are assumed to be measurable and to belong to a known parametric model class characterized by $\bm{\theta}$. The process noise~$\bm{v}_t$ and measurement noise~$\bm{w}_t$ follow the distributions~$\mathcal{V}_{\bm{\theta}}$ and $\mathcal{W}_{\bm{\theta}}$, respectively, whose characteristics (e.g., mean and covariance) may depend on $\bm{\theta}$ and whose probability density functions are denoted by $p_{\mathcal{V}}(\bm{v}_t \mid \bm{\theta})$ and $p_{\mathcal{W}}(\bm{w}_t \mid \bm{\theta})$. Conditional on~$\bm{\theta}$, they are independent and identically distributed (i.i.d.) over time, mutually independent, and independent of the initial state~$\bm{x}_0$. The assumption of a parametric model class is motivated by the fact that, in many physical systems, prior structural knowledge derived from first principles enables the formulation of expressive yet interpretable models. The same source of knowledge can be leveraged to specify prior distributions over the unknown parameters and the initial state, which motivates the following assumption.

\begin{assum}[Prior knowledge]
	\label{as:sample_from_prior}
	Prior distributions~$p(\bm{\theta})$ and $p(\bm{x}_0 \mid \bm{\theta})$ for the model parameters and the initial state are available. The unknown system parameters and the initial state are realizations drawn from these priors.
\end{assum}

In addition to the prior, a dataset~$\mathbb{D}=(\bm{u}_{0:t_0-1}, \bm{y}_{0:t_0-1})$ of past input--output measurements collected up to time~\mbox{$t_0-1$}, where $t_0 \in \mathbb{N}$, is available. Given a prediction horizon~\mbox{$H \in \mathbb{N}$} and a cost functional
\begin{equation} \label{eq:cost}
	J\left(\bm{u}_{t_0:t_0+H},\bm{x}_{t_0:t_0+H},\bm{y}_{t_0:t_0+H}\right),
\end{equation}
the objective is to determine an input sequence~$\bm{u}_{t_0:t_0+H}$ that minimizes the cost subject to two types of constraints. Input constraints, such as actuator limits, restrict the inputs to a set $\mathbb{U} \subset \mathbb{R}^{n_u}$ and can be enforced directly. Constraints on states and outputs of the form
\begin{equation} \label{eq:constraints}
	h\left(\bm{u}_{t_0:t_0+H},\bm{x}_{t_0:t_0+H},\bm{y}_{t_0:t_0+H}\right)\leq 0
\end{equation}
depend on the uncertain dynamics and cannot be guaranteed deterministically: since both parameters and disturbances may have unbounded support, this would require robustness with respect to all possible realizations of the uncertainty, which is typically infeasible or overly conservative. Instead, the objective is to ensure their satisfaction with high probability, i.e.,
\begin{equation} \label{eq:probabilistic_constraints}
	\mathbb{P}\left[ h \left(\bm{u}_{t_0:t_0+H},\bm{x}_{t_0:t_0+H},\bm{y}_{t_0:t_0+H} \right)\leq 0 \right] \geq 1-\alpha,
\end{equation}
where $\alpha \in (0,1)$ denotes the allowable risk level and the probability is taken with respect to the joint uncertainty over parameters, latent states, and future disturbances that remains given the available data~$\mathbb{D}$. Here, $h(\cdot)$ is assumed measurable and, without loss of generality, scalar-valued, since multiple constraints can be aggregated via $h = \max(h_1, \dots, h_m)$.

The formulation~\eqref{eq:system}--\eqref{eq:probabilistic_constraints} captures all four challenges identified in Section~\ref{sec:introduction} within a single problem statement, whose joint treatment is the focus of this article.

\section{Particle Markov Chain Monte Carlo Methods}
\label{sec:pmcmc}

In order to solve the optimal control problem (OCP)~\mbox{\eqref{eq:system}--\eqref{eq:probabilistic_constraints}}, it is essential to reduce uncertainty about the system by incorporating the information contained in the observed data~$\mathbb{D}$. If decisions were based solely on the prior distribution, the uncertainty over future state trajectories obtained by propagating the prior over $(\bm{\theta},\bm{x}_0)$ through the stochastic dynamics would typically be excessively large, leading either to infeasibility of the probabilistic constraints or to overly conservative solutions. Within the Bayesian framework, this reduction is achieved by inferring the posterior distribution~$p(\bm{\theta}, \bm{x}_{0:t_0-1} \mid \mathbb{D})$, which combines prior knowledge with the information provided by the measurements. For nonlinear state-space models with latent states, this posterior is analytically intractable, since its normalizing constant is an integral over the parameters and all latent state trajectories that generally admits no closed-form solution.

Approximate inference methods address this intractability in different ways. Variational approaches (e.g.,~\cite{courts2023}) approximate the posterior within a chosen parametric family, but since the resulting approximation error is generally hard to quantify, they do not readily support rigorous probabilistic guarantees. We therefore employ PMCMC methods~\cite{andrieu2010}, which embed sequential Monte Carlo algorithms, also known as particle filters, in Markov chain Monte Carlo (MCMC) samplers. Given the model parameters, a particle filter approximates the conditional distribution of latent state trajectories using weighted particles and provides an unbiased likelihood estimate, thereby enabling the construction of valid MCMC kernels targeting the exact joint posterior. The two main variants for joint inference of parameters and latent states are briefly reviewed below.

\paragraph*{Particle marginal Metropolis--Hastings (PMMH)}
At each iteration, a candidate parameter~$\bm{\theta}'$ is drawn from a proposal distribution~$q(\bm{\theta}' \mid \bm{\theta}^{(i)})$, and a particle filter with~$N$ particles computes an unbiased estimate~$\hat{p}(\mathbb{D} \mid \bm{\theta}')$ of the likelihood~$p(\mathbb{D} \mid \bm{\theta}') = p(\bm{y}_{0:t_0-1} \mid \bm{u}_{0:t_0-1}, \bm{\theta}')$. The candidate is accepted with probability
\begin{equation}\label{eq:acceptance_probability}
	\rho = \min\!\left( 1, \frac{ \hat{p}(\mathbb{D} \mid \bm{\theta}')\, p(\bm{\theta}')\, q(\bm{\theta}^{(i)} \mid \bm{\theta}') }{ \hat{p}(\mathbb{D} \mid \bm{\theta}^{(i)})\, p(\bm{\theta}^{(i)})\, q(\bm{\theta}' \mid \bm{\theta}^{(i)}) } \right),
\end{equation}
in which case~$\bm{\theta}^{(i+1)} = \bm{\theta}'$ and a state trajectory~$\bm{x}_{0:t_0-1}^{(i+1)}$ is sampled from the corresponding particle system. Otherwise, the current parameter, trajectory, and likelihood estimate are retained. The efficiency of PMMH depends on how well the proposal matches the posterior and on the number of particles~$N$, which controls the variance of the log-likelihood estimate. If this variance is high, an unusually large estimate can cause subsequent candidates to be rejected over many iterations. Both can be adapted by incorporating the data gradually and using the samples of each stage to tune the next: the proposal covariance is set to their empirical covariance, and $N$ is chosen to achieve a target variance of the log-likelihood estimate~\cite{doucet2015}. Within each stage, the proposal scale can additionally be adjusted by a \mbox{Robbins--Monro} recursion that drives the acceptance rate toward a target value~\cite{andrieu2008}.

\paragraph*{Particle Gibbs (PG)}
Instead of estimating the likelihood, PG alternates between drawing a new state trajectory with a conditional particle filter, which retains the previous trajectory as a reference particle, and sampling the parameters from~$p(\bm{\theta} \mid \bm{x}_{0:t_0-1}, \mathbb{D})$. This avoids the accept--reject step of PMMH and can improve mixing, but requires that this conditional distribution can be sampled directly, which is often achieved by choosing a conjugate prior~\cite{svensson2017}. For long data records, path degeneracy in the conditional particle filter can make consecutive trajectories highly correlated, which ancestor sampling~\cite{lindsten2014} mitigates even with few particles.

% Empty run-in paragraph heading to match the vertical spacing of the preceding \paragraph* blocks.
\paragraph*{}\leavevmode\unskip
Under mild regularity conditions, both PMMH and PG admit $p(\bm{\theta}, \bm{x}_{0:t_0-1} \mid \mathbb{D})$ as their invariant distribution and generate asymptotically exact samples for any finite number of particles~\cite{andrieu2010}, provided that any adaptation of the proposal is stopped or vanishes asymptotically~\cite{andrieu2008}. In practice, early samples are influenced by the initialization and are discarded as burn-in. Although no rigorous guarantees exist for the required length~$K_{\mathrm{b}}$, its choice is guided by empirical convergence diagnostics. Moreover, successive samples are correlated, whereas independent samples are required in the following. Two remedies are common: thinning, whereby only every $d$th sample is retained with $d$ chosen from the empirical autocorrelation function such that the remaining serial dependence is negligible, or running an independent chain for each retained sample, which yields independence by construction at the cost of a separate burn-in per chain. In either case, the retained draws can be treated as independent samples from the posterior, as formalized in the following assumption.

\begin{assum}[Independent posterior samples]
	\label{as:iid_posterior}
	The employed PMCMC procedure provides $K$ independent samples $\{(\bm{\theta}^{[k]},\bm{x}_{0:t_0-1}^{[k]})\}_{k=1}^{K}$ from $p(\bm{\theta}, \bm{x}_{0:t_0-1} \mid \mathbb{D})$.
\end{assum}

Each sample represents a realization of the unknown dynamics and latent states that is consistent with the observed data and prior knowledge. These samples serve as the basis for the control framework developed in the following sections. The framework is not tied to a specific PMCMC method, provided that the sampler satisfies Assumption~\ref{as:iid_posterior}.

\section{Validation of Control Laws}
\label{sec:validation}

For control purposes, and particularly in safety-critical applications, it is essential to assess how the uncertainty remaining after incorporating the data affects the robustness of a control law. We therefore consider the probability that the trajectories generated by a given control law satisfy the constraints~\eqref{eq:constraints} under the posterior predictive distribution induced by $p(\bm{\theta}, \bm{x}_{0:t_0-1} \mid \mathbb{D})$ and the disturbance model. Since this probability cannot be computed in closed form, we resort to randomized methods, which assess such probabilities from sampled uncertainty realizations with rigorous statistical guarantees~\cite{tempo2013,alamo2015}. These methods, however, assume that the underlying distribution can be sampled directly, which is not the case here. The PMCMC procedure of Section~\ref{sec:pmcmc} closes this gap: combined with forward propagation to the current state and sampling of future disturbances, its posterior samples yield realizations of the posterior predictive distribution, on which we build a validation procedure with finite-sample guarantees.

For the remainder of this section, we consider measurable causal control laws~$\bm{u}_t=\bm{\pi}(\bm{y}_{0:t-1}, \bm{u}_{0:t-1}, t)$ taking values in $\mathbb{U}$, which include fixed input sequences and output feedback laws. Due to the parametric state-space representation, future states and outputs are uniquely determined by the inputs together with the parameters~$\bm{\theta}$, the current state~$\bm{x}_{t_0}$, and the noise realizations. A sample~$\bm{\xi}^{[k]}$ of these exogenous quantities from $p(\bm{\theta}, \bm{x}_{t_0}, \bm{v}_{t_0:t_0+H}, \bm{w}_{t_0:t_0+H}\mid \mathbb{D})$, referred to as a \emph{scenario}, therefore fully determines a future trajectory for any given control law. Algorithm~\ref{alg:scenario_generation} generates such scenarios by drawing parameters and past state trajectories via PMCMC, sampling noise realizations from the parameterized distributions, and propagating the state forward to~$t_0$.

\begin{algorithm}[t]
	\caption{Scenario generation}
	\label{alg:scenario_generation}
	\begin{algorithmic}[1]
		\REQUIRE
			\parbox[t]{0.8\linewidth}{
				Dataset~$\mathbb{D}=(\bm{u}_{0:t_0-1}, \bm{y}_{0:t_0-1})$, \\
				model~$\{\bm{f}_{\bm{\theta}}(\cdot),\bm{g}_{\bm{\theta}}(\cdot), \mathcal{V}_{\bm{\theta}}, \mathcal{W}_{\bm{\theta}}\}$, \\
				priors~$p(\bm{\theta})$ and $p(\bm{x}_0 \mid \bm{\theta})$, \\
				horizon~$H$, number of scenarios~$K$
			}
		\ENSURE	
			\parbox[t]{0.8\linewidth}{
				Scenarios~$\{\bm{\xi}^{[k]}\}_{k=1}^{K}$, where \\
				$\bm{\xi}^{[k]}=(\bm{\theta}^{[k]},\bm{x}_{t_0}^{[k]},\bm{v}_{t_0:t_0+H}^{[k]},\bm{w}_{t_0:t_0+H}^{[k]})$
			}
		\FOR {$k = 1$ \TO $K$}
			\STATE Draw $(\bm{\theta}^{[k]}, \bm{x}_{0:t_0-1}^{[k]} ) \sim p(\bm{\theta}, \bm{x}_{0:t_0-1}\mid \mathbb{D})$ \\ using PMCMC
			\STATE Sample $\bm{v}_{t_0-1:t_0+H}^{[k]}$ with $\bm{v}_t^{[k]} \sim \mathcal{V}_{\bm{\theta}^{[k]}}$ \\ for $t=t_0-1,\ldots,t_0+H$
			\STATE Sample $\bm{w}_{t_0:t_0+H}^{[k]}$ with $\bm{w}_t^{[k]} \sim \mathcal{W}_{\bm{\theta}^{[k]}}$ \\ for $t=t_0,\ldots,t_0+H$
			\STATE $\bm{x}_{t_0}^{[k]} \gets \bm{f}_{\bm{\theta}^{[k]}}(\bm{x}_{t_0-1}^{[k]},\bm{u}_{t_0-1},\bm{v}_{t_0-1}^{[k]})$
			\STATE Store scenario~$\bm{\xi}^{[k]}$
		\ENDFOR
	\end{algorithmic}
\end{algorithm}

Statistical validation requires the scenarios to be independent of the control law, which we formalize as follows.

\begin{assum}[Independence of validation scenarios]
	\label{as:independent_control_law}
	The control law~$\bm{u}_t=\bm{\pi}(\cdot)$ is independent of the validation scenarios used to assess it, that is, it is neither designed nor adapted based on these scenarios.
\end{assum}

This assumption is not restrictive in the present setting. Although the dataset~$\mathbb{D}$ is fixed, Algorithm~\ref{alg:scenario_generation} can generate arbitrarily many independent scenarios, so that separate scenarios can be used for synthesis and validation without reducing the information available for controller design. Let $K_{\mathrm{val}}$ denote the number of validation scenarios and $\mathcal{S}_{\mathrm{val}}$ their index set.

To assess the robustness of a control law, the closed-loop system is simulated over the horizon~$H$ for each validation scenario~$\bm{\xi}^{[k]}$, $k \in \mathcal{S}_{\mathrm{val}}$, and it is checked whether the constraints~\eqref{eq:constraints} are satisfied (Algorithm~\ref{alg:forward_simulation}). The constraint value
\begin{equation}
	Z^{[k]} = h \left( \bm{u}_{t_0:t_0+H}^{[k]},\bm{x}_{t_0:t_0+H}^{[k]},\bm{y}_{t_0:t_0+H}^{[k]} \right)
\end{equation}
indicates a violation whenever $Z^{[k]} > 0$, and the total number of violations over the validation set is
\begin{equation}
	V = \sum_{k \in \mathcal{S}_{\mathrm{val}}} \mathbf{1}\left\{Z^{[k]} > 0\right\},
\end{equation}
where $\mathbf{1}\{\cdot\}$ denotes the indicator function. Based on this count, the following theorem provides a finite-sample guarantee for constraint satisfaction.

\begin{algorithm}[t]
	\caption{Validation of a control law}
	\label{alg:forward_simulation}
	\begin{algorithmic}[1]
		\REQUIRE \parbox[t]{0.8\linewidth}{
			Validation scenarios~$\{\bm{\xi}^{[k]}\}_{k \in \mathcal{S}_{\mathrm{val}}}$, \\
			control law~$\bm{u}_t=\bm{\pi}(\bm{y}_{0:t-1},\bm{u}_{0:t-1},t)$, \\
			horizon~$H$
		}
		\ENSURE Number of scenarios with constraint violations~$V$
		\STATE $V \gets 0$
		\FOR{$k \in \mathcal{S}_{\mathrm{val}}$}
			\STATE $(\bm{u}_{0:t_0-1}^{[k]}, \bm{y}_{0:t_0-1}^{[k]}) \gets (\bm{u}_{0:t_0-1}, \bm{y}_{0:t_0-1})$
			\FOR {$t = t_0$ \TO $t_0+H$}
				\STATE $\bm{u}_t^{[k]} \gets \bm{\pi}(\bm{y}_{0:t-1}^{[k]},\bm{u}_{0:t-1}^{[k]},t)$
				\STATE $\bm{x}_{t+1}^{[k]} \gets \bm{f}_{\bm{\theta}^{[k]}}(\bm{x}_t^{[k]} ,\bm{u}_t^{[k]},\bm{v}_t^{[k]})$
				\STATE $\bm{y}_{t}^{[k]} \gets \bm{g}_{\bm{\theta}^{[k]}}( \bm{x}_{t}^{[k]} ,\bm{u}_{t}^{[k]})+\bm{w}_t^{[k]}$
			\ENDFOR
			\IF{$h(\bm{u}_{t_0:t_0+H}^{[k]},\bm{x}_{t_0:t_0+H}^{[k]},\bm{y}_{t_0:t_0+H}^{[k]}) > 0$}
				\STATE $V \gets V + 1$
			\ENDIF
		\ENDFOR
	\end{algorithmic}
\end{algorithm}

\begin{thm}[Robustness guarantee]
	\label{theo:robustness_guarantees}
	Suppose that Assumptions~\ref{as:sample_from_prior},~\ref{as:iid_posterior}, and~\ref{as:independent_control_law} hold. Let $V$ denote the number of constraint violations observed among the $K_{\mathrm{val}}$ validation scenarios, let $1-\beta$ with $\beta \in (0,1)$ be a chosen confidence level, and let $\nu := \mathbb{P}\left(Z > 0 \mid \mathbb{D}, \bm{\pi}\right)$ denote the violation probability under the posterior predictive distribution. If $V < K_{\mathrm{val}}$, let $\delta \in (0,1)$ be the unique solution of
	\begin{equation}\label{eq:clopper-pearson}
		\sum_{i=0}^V \binom{K_{\mathrm{val}}}{i}\delta^i(1-\delta)^{K_{\mathrm{val}}-i}=\beta;
	\end{equation}
	if $V = K_{\mathrm{val}}$, set $\delta = 1$. Then, with confidence at least $1-\beta$, it holds that $\nu \le \delta$, i.e., the constraints~\eqref{eq:constraints} are satisfied under~$\bm{\pi}(\cdot)$ with posterior predictive probability at least $1-\delta$.
\end{thm}
\begin{pf}
	By Assumption~\ref{as:sample_from_prior}, the true system is a realization of the probabilistic model, so that the posterior predictive distribution describes its possible future trajectories given~$\mathbb{D}$. By Assumptions~\ref{as:iid_posterior} and~\ref{as:independent_control_law}, the validation scenarios are independent draws from this distribution and independent of~$\bm{\pi}(\cdot)$. Conditional on $\mathbb{D}$ and $\bm{\pi}$, the violation indicators are therefore i.i.d.\ Bernoulli with parameter~$\nu$, and $V \sim \mathrm{Binomial}(K_{\mathrm{val}},\nu)$. For this binomial model, the Clopper--Pearson construction~\cite{clopper1934} yields the upper $(1-\beta)$ confidence bound~$\delta$ on $\nu$ given by~\eqref{eq:clopper-pearson} for $V < K_{\mathrm{val}}$, while $\nu \le \delta = 1$ holds trivially for $V = K_{\mathrm{val}}$. \hfill$\blacksquare$
\end{pf}

For a fixed empirical violation rate~$V/K_{\mathrm{val}}$, the guarantee~$1-\delta$ strengthens as the number of validation scenarios~$K_{\mathrm{val}}$ grows, and for a fixed $K_{\mathrm{val}}$, fewer violations yield a stronger guarantee~\cite{clopper1934}. The confidence parameter~$\beta$ controls the statistical reliability of the bound: if the validation procedure were repeated many times, the bound~$1-\delta$ would hold in at least a $1-\beta$ fraction of these repetitions. Since $\delta$ is a Beta-distribution quantile that changes only slowly in~$\beta$, the confidence parameter can be chosen very small (e.g., $\beta = 10^{-3}$) without significantly weakening the guarantee~\cite{thulin2014}. If several candidates are validated, $\beta$ must be allocated across attempts to account for multiple testing~\cite{hochberg1987}.

\section{Risk-Sensitive Optimal Control}
\label{sec:optimal_control}

The validation procedure of the previous section certifies a given control law but does not construct one. This section develops a systematic method that computes and certifies an input sequence for the OCP~\eqref{eq:system}--\eqref{eq:probabilistic_constraints}. As in sequential randomized methods~\cite{tempo2013,alamo2015}, a candidate is synthesized on $K_{\mathrm{train}}$ training scenarios with index set~$\mathcal{S}_{\mathrm{train}}$ and then assessed on an independent validation set of size~$K_{\mathrm{val}}$. Section~\ref{subsec:scenario_synthesis} formulates a synthesis problem whose conservativeness is adjusted by a single risk weight, Section~\ref{subsec:adaptive_calibration} describes how this weight is calibrated across repeated synthesis--validation attempts, and Section~\ref{subsec:oc_analysis} analyzes the resulting procedure.

\subsection{CVaR-Penalized Scenario Program}
\label{subsec:scenario_synthesis}

Given the training scenarios~$\mathcal{S}_{\mathrm{train}}$, we seek a candidate input sequence that approximates the solution of the stochastic OCP~\eqref{eq:system}--\eqref{eq:probabilistic_constraints}. To this end, we formulate a scenario-based optimization problem in which the training scenarios are treated as deterministic realizations of the uncertainty. A natural approach would be to enforce the constraints for all scenarios, but this is overly conservative, since a few low-probability scenarios can severely restrict the feasible set. Explicitly selecting which scenarios may violate the constraints, on the other hand, leads to an expensive combinatorial problem. Instead, we penalize tail risk through the CVaR, which allows the conservativeness to be adjusted continuously without changing the structure of the problem.

The CVaR quantifies the expected value of the upper tail of a distribution. The constraint value~$Z = h(\cdot)$ and the cost~$J$ are random variables due to the uncertain dynamics and disturbances, assumed to be integrable with respect to the posterior predictive distribution for every admissible input sequence. Minimizing the CVaR of~$Z$ tends to reduce both the magnitude and the frequency of violations, and since fewer violations yield a tighter bound in Theorem~\ref{theo:robustness_guarantees}, this improves the certifiable robustness of the synthesized input sequence. For a tail level~$\tau \in (0,1)$, the CVaR is defined as
\begin{equation}\label{eq:cvar}
	\mathrm{CVaR}_{\tau}(Z) = \min_{\eta \in \mathbb{R}} \left( \eta + \frac{1}{\tau}\,\mathbb{E}\left[(Z-\eta)_+\right] \right),
\end{equation}
where $(a)_+ = \max(a,0)$ and the smallest minimizer is the value-at-risk (VaR)
\begin{equation}\label{eq:var}
	\mathrm{VaR}_{\tau}(Z) = \inf\{z \in \mathbb{R} : \mathbb{P}(Z > z) \leq \tau\},
\end{equation}
i.e., the $(1-\tau)$-quantile of~$Z$~\cite{rockafellar2002}. We set $\tau = \alpha$ by default, which aligns the penalized tail with the desired robustness level. Combining the expected cost with a penalty on the CVaR yields the risk-sensitive objective
\begin{equation}\label{eq:population_objective}
	\min_{\bm{u}_{t_0:t_0+H} \in \mathcal{U}} \; (1-\omega)\,\mathbb{E}[J] \;+\; \omega\, \mathrm{CVaR}_{\tau}(Z),
\end{equation}
where $\mathcal{U} = \mathbb{U}^{H+1}$ denotes the set of admissible input sequences. The risk weight~$\omega \in [0,1]$ balances nominal performance and robustness: at $\omega = 0$, the CVaR term is inactive and~\eqref{eq:population_objective} reduces to the risk-neutral problem of minimizing the expected cost, whereas at $\omega = 1$ the cost is disregarded and only the tail risk is minimized.

Since the expectations in~\eqref{eq:population_objective} are generally intractable, we replace them by sample averages over the training scenarios, each of which induces realizations of the cost~$J^{[k]}$ and the constraint value~$Z^{[k]}$. This yields the scenario-based optimization problem
\begin{subequations} \label{eq:scenario_OCP}
	\begin{align}
		\min_{\bm{u}_{t_0:t_0+H} \in \mathcal{U},\eta} &\; \frac{1-\omega}{K_{\mathrm{train}}}\sum_{k \in \mathcal{S}_{\mathrm{train}}} J^{[k]} \nonumber\\
		&+ \omega \left( \eta + \frac{1}{\tau K_{\mathrm{train}}} \sum_{k \in \mathcal{S}_{\mathrm{train}}} \left(Z^{[k]}-\eta \right)_+ \right) \label{eq:OCP_objective} \\
		\text{s.t.} \quad
		\forall k \in & \mathcal S_{\mathrm{train}}, \quad
		\forall t\in \left\{t_0,\dots,t_0+H \right\} \nonumber\\
		\bm{x}_{t+1}^{[k]} &=
		\bm{f}_{\bm{\theta}^{[k]}}
		\left(\bm{x}_t^{[k]},\bm{u}_t,\bm{v}_t^{[k]} \right),
		\label{eq:OCP_dynamic_constraints_x}\\
		\bm{y}_{t}^{[k]} &=
		\bm{g}_{\bm{\theta}^{[k]}}
		\left(\bm{x}_{t}^{[k]},\bm{u}_{t} \right)
		+\bm{w}_t^{[k]},
		\label{eq:OCP_dynamic_constraints_y}\\
		J^{[k]} &=
		J\left(\bm{u}_{t_0:t_0+H},\bm{x}_{t_0:t_0+H}^{[k]},\bm{y}_{t_0:t_0+H}^{[k]}\right),
		\label{eq:OCP_J}\\
		Z^{[k]} &=
		h\left(\bm{u}_{t_0:t_0+H},\bm{x}_{t_0:t_0+H}^{[k]},\bm{y}_{t_0:t_0+H}^{[k]}\right).
		\label{eq:OCP_H}
	\end{align}
\end{subequations}
Since the only hard constraints apart from the dynamics are the input constraints, while violations of~\eqref{eq:constraints} enter via the CVaR penalty, problem~\eqref{eq:scenario_OCP} is feasible for any~$\omega$. We assume that a solution can be computed by a suitable numerical optimization method~\cite{nocedal2006}, which may require additional regularity of the dynamics, cost, and constraint functions. Since the problem is generally non-convex, global optimality cannot be guaranteed. The certificate, however, does not rely on optimality, and locally optimal solutions are often sufficient in practice. Importantly, the training and validation set sizes can be chosen independently: while $K_{\mathrm{train}}$ is limited by the computational cost of solving~\eqref{eq:scenario_OCP}, validation only requires forward simulation, so that $K_{\mathrm{val}}$ can be chosen substantially larger, yielding a tighter bound in Theorem~\ref{theo:robustness_guarantees}.

\subsection{Adaptive Calibration of the Risk Weight}
\label{subsec:adaptive_calibration}

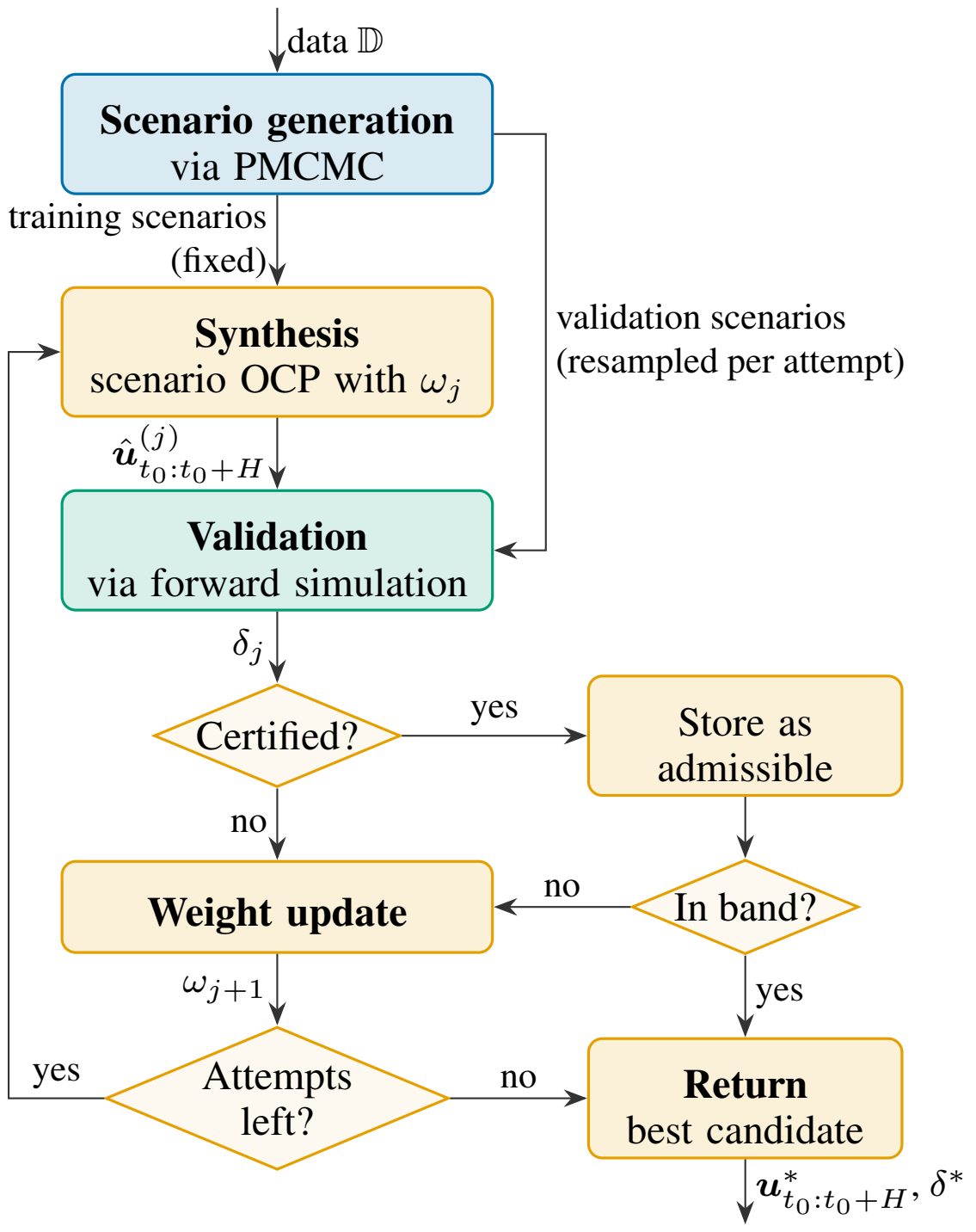
\begin{figure}[t]
	\centering
	\begin{tikzpicture}[
		font=\footnotesize,
		>={Stealth[length=1.8mm]},
		node distance=5.5mm,
		box/.style={draw=#1, fill=#1!15, line width=0.6pt, rounded corners=3pt, align=center, text width=3.1cm, inner sep=3pt, minimum height=7mm},
		dec/.style={draw=oiOrange, fill=oiOrange!6, line width=0.6pt, diamond, aspect=2.4, align=center, inner sep=0.5pt},
		lbl/.style={font=\scriptsize, inner sep=2pt},
		flow/.style={->, draw=gray!40!black},
		]
		
		\node[box=oiBlue] (scen) {\textbf{Scenario generation}\\ via PMCMC};
		\node[box=oiOrange, below=7mm of scen] (ocp) {\textbf{Synthesis}\\ scenario OCP with $\omega_j$};
		\node[box=oiGreen, below=of ocp] (valid) {\textbf{Validation}\\ via forward simulation};
		\node[dec, below=of valid] (cert) {Certified?};
		\node[box=oiOrange, below=of cert] (update) {\textbf{Weight update}};
		\node[dec, below=of update] (budget) {Attempts\\ left?};
		
		\node[box=oiOrange, text width=2.2cm] (adm) at ([xshift=3.6cm]cert.center) {Store as admissible};
		\node[dec] (band) at ([xshift=3.6cm]update.center) {In band?};
		\node[box=oiOrange, text width=2.2cm] (ret) at ([xshift=3.6cm]budget.center) {\textbf{Return}\\ best candidate};
		
		\draw[flow, <-] (scen.north) -- node[lbl, right] {data $\mathbb{D}$} ++(0,5mm);
		\draw[flow] (ret.south) -- node[lbl, right] {$\bm{u}^*_{t_0:t_0+H}$, $\delta^*$} ++(0,-5mm);
		
		\draw[flow] (scen) -- node[lbl, left, align=right] {training scenarios\\ (fixed)} (ocp);
		\draw[flow] (scen.east) -- ++(4mm,0) |- node[lbl, right, align=left, pos=0.25] {validation scenarios\\ (resampled per attempt)} (valid.east);
		\draw[flow] (ocp) -- node[lbl, left] {$\hat{\bm{u}}^{(j)}_{t_0:t_0+H}$} (valid);
		\draw[flow] (valid) -- node[lbl, left] {$\delta_j$} (cert);
		\draw[flow] (cert) -- node[lbl, above] {yes} (adm);
		\draw[flow] (cert) -- node[lbl, left] {no} (update);
		\draw[flow] (adm) -- (band);
		\draw[flow] (band) -- node[lbl, above] {no} (update);
		\draw[flow] (band) -- node[lbl, right] {yes} (ret);
		\draw[flow] (update) -- node[lbl, left] {$\omega_{j+1}$} (budget);
		\draw[flow] (budget) -- node[lbl, above] {no} (ret);
		\coordinate (lx) at ([xshift=-4mm]ocp.west);
		\draw[flow] (budget.west) -- node[lbl, above] {yes} (budget.west -| lx) |- (ocp.west);
	\end{tikzpicture}
	
	\caption{Overview of the proposed approach. Training and validation scenarios are generated from PMCMC posterior samples. In attempt~$j$, a candidate input sequence is synthesized from the training scenarios with risk weight~$\omega_j$ and evaluated on a fresh validation set, yielding a high-confidence bound~$\delta_j$ on its violation probability. The risk weight is updated until a candidate is certified within the acceptance band or the attempt budget is exhausted. Colors indicate scenario generation (blue, Section~\ref{sec:pmcmc} and Algorithm~\ref{alg:scenario_generation}), validation (green, Section~\ref{sec:validation}), and synthesis and calibration (orange, Section~\ref{sec:optimal_control}).}
	\label{fig:calibration}
\end{figure}

Since the appropriate risk weight~$\omega$ is not known a priori, it is calibrated in a synthesis--validation loop, illustrated in Figure~\ref{fig:calibration}. In attempt~$j$, the scenario OCP~\eqref{eq:scenario_OCP} is solved for a given risk weight~$\omega_j$, and the resulting candidate input sequence~$\hat{\bm{u}}^{(j)}_{t_0:t_0+H}$ is evaluated on an independent validation set, yielding the bound~$\delta_j$ of Theorem~\ref{theo:robustness_guarantees}. We call the candidate certified if $\delta_j \leq \alpha$. The search terminates once a certified candidate is found whose bound lies within a specified tolerance of~$\alpha$. Otherwise, $\omega$ is adjusted and the synthesis step is repeated. The training scenarios are reused across attempts to keep the approximation of the objective consistent, while a fresh validation set is generated at each attempt to maintain independence (Assumption~\ref{as:independent_control_law}).

Each attempt, however, constitutes a separate statistical test, and repeated validation inflates the probability of a spurious certification. To maintain the overall confidence level~$1-\beta$, at most $n_{\mathrm{max}}$ attempts are permitted and the confidence parameter is split uniformly as $\beta_{\mathrm{att}} = \beta / n_{\mathrm{max}}$~\cite{hochberg1987}, which weakens the individual bounds only marginally (cf.~Section~\ref{sec:validation}). This limited budget requires a schedule that explores the range of~$\omega$ efficiently and reaches the most conservative choice~$\omega = 1$ by the final attempt.

\subsubsection{Response model}
To select the risk weight efficiently, we characterize how the probability~$\nu$ that a scenario violates the constraints (cf.~Theorem~\ref{theo:robustness_guarantees}) depends on it. For $\omega < 1$, dividing~\eqref{eq:population_objective} by $1-\omega$ yields the equivalent objective
\begin{equation}\label{eq:penalty_objective} 
	\min_{\bm{u}_{t_0:t_0+H} \in \mathcal{U}} \; \mathbb{E}[J] \;+\; \lambda\, \mathrm{CVaR}_{\tau}(Z), \qquad \lambda = \frac{\omega}{1-\omega},
\end{equation}
so that the trade-off is governed by the relative penalty weight~$\lambda \in [0,\infty)$, which provides a natural parameterization for the calibration. The violation probability~$\nu$ typically exhibits a saturating response: for small $\lambda$, the CVaR term has little influence on the solution, whereas for large $\lambda$, further increasing its weight produces diminishing changes and $\nu$ may approach a floor induced by the remaining uncertainty. The calibration therefore targets the transition region in between, where we locally linearize $\log\nu$ in $\log\lambda$, which corresponds to a power law~$\nu \approx c\lambda^s$ and yields
\begin{equation}\label{eq:power_law} 
	\log \nu \approx a + s\log\lambda, \qquad s=\frac{\mathrm d\log\nu}{\mathrm d\log\lambda}<0.
\end{equation}
The local elasticity~$s$ is assumed to be negative, since a larger weight on the tail risk tends to shift the solution toward fewer violations (cf.~Section~\ref{subsec:oc_analysis}).

Since $s$ is unknown, it is estimated from the preceding attempts, each of which yields an empirical violation rate~\mbox{$\hat{\nu}_j = V_j / K_{\mathrm{val}}$}, a noisy estimate of~$\nu_j$. To balance locality against noise averaging, we use exponentially weighted least squares in $(\log\lambda, \log\hat{\nu})$ and exclude saturated attempts with $V_j \in \{0, K_{\mathrm{val}}\}$, which carry no slope information. Until sufficiently many attempts are available, an initial value~$s_{\mathrm{init}} < 0$ is used, and $\hat{s}_j$ is bounded from above by~$s_{\mathrm{max}} < 0$ to preserve a decreasing response under sampling noise.

\subsubsection{Weight update}
The next weight is chosen to reach a target violation probability~$\nu_{\mathrm{tar}}$. Since the bound~$\delta$ is strictly increasing in the violation count~$V$, certification at level~$1-\alpha$ is equivalent to $V_j \leq V_\alpha^*$, where
\begin{equation}\label{eq:V_alpha_star} 
	V_\alpha^* = \max\left\{v \in \mathbb{N}^0 : \delta_v \leq \alpha\right\} 
\end{equation}
is the largest violation count that still permits certification, and $\delta_v$ follows from~\eqref{eq:clopper-pearson} with $V = v$ and $\beta = \beta_{\mathrm{att}}$. We assume that $K_{\mathrm{val}}$ is large enough that $V_\alpha^* \geq 1$, since otherwise certification would require zero violations and the target would degenerate. Targeting the corresponding rate~$V_\alpha^*/K_{\mathrm{val}}$ directly would place $\nu_{\mathrm{tar}}$ exactly at the certification boundary, where sampling variability in the validation set alone would lead to frequent rejections. We therefore set $\nu_{\mathrm{tar}} = (1-m) V_\alpha^*/K_{\mathrm{val}}$ with a relative margin~$m \in [0,1)$. Together with~$\hat{s}_j$ and the current outcome,~\eqref{eq:power_law} then yields the step
\begin{equation}\label{eq:weight_update} 
	\Delta_j = \frac{\log\nu_{\mathrm{tar}} - \log\hat{\nu}_j}{\hat{s}_j}, \quad \log\lambda_{j+1} = \log\lambda_j + \Delta_j,
\end{equation} 
which increases~$\lambda$ when the observed rate exceeds the target and decreases it otherwise, with a magnitude that adapts to the local sensitivity. Since~\eqref{eq:power_law} is only a local approximation with a noisy slope estimate, $|\Delta_j|$ is bounded in practice, which also covers the case $V_j = 0$, and $\lambda_{j+1}$ is restricted to lie between the largest risk weight that failed to certify and the smallest one that certified with more margin than required. These choices affect only the number of attempts, not what is certified.

Since the update~\eqref{eq:weight_update} ignores the remaining budget, an inaccurate slope estimate may exhaust all attempts without certification, even if a certifiable candidate exists at a larger risk weight. As a conservative certified solution is preferable to none, a recovery phase is entered over the final $n_{\mathrm{rec}}$ attempts if no candidate has certified yet. With $r_j = n_{\mathrm{max}} - j$ remaining attempts, the recovery proposal
\begin{equation}\label{eq:recovery_step}
	\omega_{j+1}^{\mathrm{rec}} = \omega_j + \frac{1 - \omega_j}{r_j}
\end{equation}
advances~$\omega$ in equal steps and reaches $\omega = 1$, which corresponds to $\lambda = \infty$, exactly at the final attempt. During recovery, the more conservative of the two proposals is used,
\begin{equation}\label{eq:combined_update}
	\omega_{j+1} = \max\left\{\frac{\lambda_{j+1}}{1+\lambda_{j+1}},\; \omega_{j+1}^{\mathrm{rec}}\right\},
\end{equation}
and once a candidate has certified, recovery is disabled.

\subsubsection{Termination}
Every certified candidate is admissible, regardless of its risk weight. However, a bound~$\delta_j$ well below~$\alpha$ typically indicates an unnecessarily large~$\lambda$ at the expense of nominal performance. The search therefore terminates once the bound lies within the acceptance band
\begin{equation}\label{eq:acceptance_band}
	\alpha - \alpha_{\mathrm{tol}} \;\leq\; \delta_j \;\leq\; \alpha
\end{equation}
with tolerance~$\alpha_{\mathrm{tol}} \in (0,\alpha]$, or once the attempt budget is exhausted. The admissible candidate with the smallest empirical cost
\begin{equation}\label{eq:empirical_cost}
	\hat{J} = \frac{1}{K_{\mathrm{val}}} \sum_{k \in \mathcal{S}_{\mathrm{val}}} J\big(\bm{u}_{t_0:t_0+H}^{[k]}, \bm{x}_{t_0:t_0+H}^{[k]}, \bm{y}_{t_0:t_0+H}^{[k]}\big),
\end{equation}
evaluated along the trajectories simulated in Algorithm~\ref{alg:forward_simulation}, is then returned as~$\bm{u}^*_{t_0:t_0+H}$ together with its bound~$\delta^*$. No further correction is required, since the union bound underlying~$\beta_{\mathrm{att}}$ covers all attempts simultaneously, so the certificate holds regardless of which admissible candidate is selected. If no candidate has certified despite recovery, the candidate with the smallest bound is returned instead, with~$\delta^* = \min_j \delta_j > \alpha$ and a failure flag.

\subsection{Analysis}
\label{subsec:oc_analysis}

The validity of the certificate returned by the synthesis--validation procedure follows directly from the results of Section~\ref{sec:validation}. Each attempt is evaluated on a fresh, independent validation set, so that Theorem~\ref{theo:robustness_guarantees} applies with confidence parameter~$\beta_{\mathrm{att}}$, and a union bound over all $n_{\mathrm{max}}$ attempts yields confidence at least $1-\beta$ for the returned bound~$\delta^*$. This guarantee holds for any candidate, regardless of how it was obtained. In particular, it remains valid if the sample-average approximation is inaccurate, the synthesized solution is only locally or approximately optimal, or the local model~\eqref{eq:power_law} is a poor fit.

It remains to analyze whether the procedure finds a certified candidate. To this end, we show that the risk weight orders the synthesized solutions by tail risk, that a negative tail risk bounds the violation probability strictly below the target level, and that this yields a lower bound on the certification probability. To study these relations independently of finite-sample effects in the training scenarios, we analyze the population-level objective~\eqref{eq:population_objective}, whose minimizers are defined directly with respect to the posterior predictive distribution. In the following, we write $\bm{u} = \bm{u}_{t_0:t_0+H} \in \mathcal{U}$ for brevity and assume that this objective attains a finite minimum for every $\omega \in [0,1]$, which holds, e.g., if $\mathcal{U}$ is compact and $\mathbb{E}[J]$ and $\mathrm{CVaR}_{\alpha}(Z)$ are continuous on $\mathcal{U}$.

\begin{lem}[Monotonicity]\label{lem:cvar_monotone}
	For each $\omega \in [0,1]$, let $\bm{u}^*(\omega)$ be any global minimizer over $\mathcal{U}$ of the population objective~\eqref{eq:population_objective} with $\tau=\alpha$. Then, for all $0 \leq \omega_1 < \omega_2 \leq 1$, it holds that
	\begin{align}
		\mathrm{CVaR}_{\alpha}\left(Z(\bm{u}^*(\omega_2))\right) &\leq \mathrm{CVaR}_{\alpha}\left(Z(\bm{u}^*(\omega_1))\right), \label{eq:cvar_ordering}\\
		\mathbb{E}\left[J(\bm{u}^*(\omega_2))\right] &\geq \mathbb{E}\left[J(\bm{u}^*(\omega_1))\right]. \label{eq:cost_ordering}
	\end{align}
\end{lem}
\begin{pf}
	The result follows from standard properties of weighted-sum scalarizations in bi-objective optimization; see, e.g.,~\cite[Section~4.7.5]{boyd2004}. Define $C(\bm{u}) = \mathbb{E}[J(\bm{u})]$ and $R(\bm{u}) = \mathrm{CVaR}_{\alpha}(Z(\bm{u}))$, and consider first $\omega_2 < 1$. For $i \in \{1,2\}$, let $\bm{u}_i = \bm{u}^*(\omega_i)$ and $\lambda_i = \omega_i/(1-\omega_i)$. Since $\bm{u}_i$ minimizes~\eqref{eq:penalty_objective} with $\lambda = \lambda_i$,
	\begin{align}
		C(\bm{u}_2) + \lambda_2 R(\bm{u}_2) &\leq C(\bm{u}_1) + \lambda_2 R(\bm{u}_1), \label{eq:opt_u2} \\
		C(\bm{u}_1) + \lambda_1 R(\bm{u}_1) &\leq C(\bm{u}_2) + \lambda_1 R(\bm{u}_2). \label{eq:opt_u1}
	\end{align}
	Adding the inequalities yields $(\lambda_2 - \lambda_1)(R(\bm{u}_2) - R(\bm{u}_1)) \leq 0$. Since $\omega \mapsto \lambda$ is strictly increasing on $[0,1)$, $\lambda_2 > \lambda_1$ and hence $R(\bm{u}_2) \leq R(\bm{u}_1)$. For $\omega_2 = 1$, the objective~\eqref{eq:population_objective} reduces to $R$ alone, so that $R(\bm{u}_2) \leq R(\bm{u})$ for every $\bm{u} \in \mathcal{U}$, which extends~\eqref{eq:cvar_ordering} to the entire interval~$[0,1]$. In both cases, rearranging~\eqref{eq:opt_u1}, which relies only on the optimality of $\bm{u}_1$, and using $\lambda_1 \geq 0$ together with~\eqref{eq:cvar_ordering} gives $C(\bm{u}_1) - C(\bm{u}_2) \leq \lambda_1\left(R(\bm{u}_2) - R(\bm{u}_1)\right) \leq 0$, which proves~\eqref{eq:cost_ordering}. \hfill$\blacksquare$
\end{pf}

Lemma~\ref{lem:cvar_monotone} establishes the qualitative premise of the calibration loop: increasing $\omega$ never yields a global solution with a larger CVaR of the constraint value or a smaller expected cost. Since certification depends on the violation probability rather than on the CVaR, we connect the two under the following mild regularity condition.

\begin{assum}[Uniform lower-tail bound]\label{as:constraint_lower_bound}
	There exists $M>0$ such that
	\begin{equation}\label{eq:var_bound}
		\mathrm{VaR}_{\alpha}\left(Z(\bm{u})\right)\geq-M
	\end{equation}
	for every $\bm{u} \in \mathcal{U}$, with $\mathrm{VaR}_{\alpha}$ as defined in~\eqref{eq:var}.
\end{assum}

Assumption~\ref{as:constraint_lower_bound} is implied by a uniform lower bound on the constraint value itself, which holds whenever the admissible trajectories remain within a bounded operating region and the constraint function is continuous, as is common for physical systems. The constant~$M$ enters only the analysis and is neither required nor estimated by the procedure.

\begin{lem}[Violation-probability margin]\label{lem:margin}
	Let Assumption~\ref{as:constraint_lower_bound} hold and let $\bm{u} \in \mathcal{U}$. If $\mathrm{CVaR}_{\alpha}\left(Z(\bm{u})\right)\leq-\kappa$ for some $\kappa>0$, then
	\begin{equation}\label{eq:nu_bar}
		\mathbb{P}\left(Z(\bm{u})>0\right) \leq \alpha\left(1-\frac{\kappa}{M}\right).
	\end{equation}
\end{lem}
\begin{pf}
	Write $Z=Z(\bm{u})$, let $\nu=\mathbb{P}(Z>0)$, and recall the quantile representation~\cite{rockafellar2002}
	\begin{equation}
		\mathrm{CVaR}_{\alpha}(Z) = \frac{1}{\alpha}\int_0^\alpha \mathrm{VaR}_q(Z)\,\mathrm{d}q .
	\end{equation}
	Since $\mathrm{VaR}_q(Z)$ as defined in~\eqref{eq:var} is non-increasing in $q$, Assumption~\ref{as:constraint_lower_bound} implies $\mathrm{VaR}_q(Z)\geq-M$ for all $q\leq\alpha$. Moreover, for every $q<\nu$, $\mathbb{P}(Z>0)>q$ implies $\mathrm{VaR}_q(Z)\geq0$. It follows that $\nu<\alpha$, as otherwise the latter bound would hold almost everywhere on the integration range and yield $\mathrm{CVaR}_{\alpha}(Z)\geq0$, contradicting $\mathrm{CVaR}_{\alpha}(Z)\leq-\kappa<0$. Splitting the quantile integral at $\nu$ then gives
	\begin{equation}
		\mathrm{CVaR}_{\alpha}(Z) \geq \frac{1}{\alpha}\left(\int_0^\nu 0\,\mathrm{d}q + \int_\nu^\alpha -M\,\mathrm{d}q\right) = -\frac{\alpha-\nu}{\alpha}M .
	\end{equation}
	Combining this with $\mathrm{CVaR}_{\alpha}(Z)\leq-\kappa$ yields~\eqref{eq:nu_bar}. \hfill$\blacksquare$
\end{pf}

That a non-positive CVaR implies satisfaction of the corresponding chance constraint is well known~\cite{rockafellar2002}. Lemma~\ref{lem:margin} quantifies this implication: once the CVaR is driven below zero, the bound on the violation probability decreases linearly with the margin~$\kappa$. To apply this along the family of global solutions, let $\mathcal{U}^*(\omega)$ denote the set of global minimizers of~\eqref{eq:population_objective} over $\mathcal{U}$ at $\omega$ and define the guaranteed CVaR margin
\begin{equation}\label{eq:kappa_omega}
	\kappa(\omega) = -\sup_{\bm{u}\in\mathcal{U}^*(\omega)} \mathrm{CVaR}_{\alpha}\left(Z(\bm{u})\right).
\end{equation}
Every global minimizer at $\omega$ then has violation probability at most
\begin{equation}\label{eq:nu_omega}
	\bar{\nu}(\omega)
	=
	\begin{cases}
		\displaystyle \alpha\left(1-\frac{\kappa(\omega)}{M}\right), & \kappa(\omega)\geq0, \\
		1, & \kappa(\omega)<0 ,
	\end{cases}
\end{equation}
where the first case follows from Lemma~\ref{lem:margin} for $\kappa(\omega)>0$ and from the classical implication for $\kappa(\omega)=0$, while for $\kappa(\omega)<0$ only the trivial bound holds. An independent validation attempt therefore certifies with probability at least
\begin{equation}\label{eq:p_cert}
	p_{\mathrm{cert}}(\omega) = \sum_{v=0}^{V_\alpha^*} \binom{K_{\mathrm{val}}}{v} \bar{\nu}(\omega)^{\,v} \left(1-\bar{\nu}(\omega)\right)^{K_{\mathrm{val}}-v}.
\end{equation}
Since Lemma~\ref{lem:cvar_monotone} holds for every choice of global minimizers, $\kappa(\omega)$ is non-decreasing and hence $\bar{\nu}(\omega)$ non-increasing in $\omega$. As~\eqref{eq:p_cert} is the binomial cumulative distribution function with success probability~$\bar{\nu}(\omega)$ evaluated at $V_\alpha^*$, which decreases in the success probability, $p_{\mathrm{cert}}(\omega)$ is non-decreasing in $\omega$. This establishes the mechanism underlying the calibration: the larger the risk weight~$\omega$, the higher the guaranteed probability that the resulting candidate is certified.

The procedure exploits this in two ways. As long as no candidate has certified, the observed violation rate lies above the target, so that the update~\eqref{eq:weight_update} raises the risk weight, at a pace governed by the local model~\eqref{eq:power_law}. The recovery phase, in turn, enforces progress toward $\omega=1$ independently of the accuracy of this model. The following theorem quantifies the resulting probability of successful certification.

\begin{thm}[Certification probability]\label{theo:algorithm_success}
	Suppose Assumptions~\ref{as:sample_from_prior}--\ref{as:constraint_lower_bound} hold, and that each synthesis step returns a global minimizer of the population objective~\eqref{eq:population_objective} with $\tau=\alpha$. Then, with respect to the validation scenarios generated during calibration, the synthesis--validation procedure of Section~\ref{subsec:adaptive_calibration} returns a certified input sequence with probability at least
	\begin{equation}\label{eq:success_probability}
		1- \prod_{k=1}^{n_{\mathrm{rec}}} \left[ 1-p_{\mathrm{cert}}\left(\frac{k}{n_{\mathrm{rec}}}\right) \right] \geq p_{\mathrm{cert}}(1).
	\end{equation}
\end{thm}
\begin{pf}
	Since the risk weights depend on the preceding validation outcomes, we restrict attention to the recovery phase, which is conservative, as certification at an earlier attempt can only increase the overall success probability. Let $\omega_{\mathrm{e}}$ denote the risk weight at the entry to recovery and let $\widetilde{\omega}_k$ denote the risk weight of its $k$th subsequent attempt. Conditional on no previous certification, the recovery phase remains active, and~\eqref{eq:recovery_step} together with~\eqref{eq:combined_update} yields
	\begin{equation}\label{eq:recovery_lower_bound}
		\widetilde{\omega}_k \geq \omega_{\mathrm{e}} + \frac{k}{n_{\mathrm{rec}}}\left(1-\omega_{\mathrm{e}}\right) \geq \frac{k}{n_{\mathrm{rec}}}, \quad k=1,\ldots,n_{\mathrm{rec}} ,
	\end{equation}
	where the second inequality removes the dependence on the preceding attempts. Now condition on any history under which no earlier attempt has certified. This history determines both $\widetilde{\omega}_k$ and the synthesized candidate. Since the validation scenarios are independent of that candidate by Assumption~\ref{as:independent_control_law}, the conditional probability of certification is at least $p_{\mathrm{cert}}(\widetilde{\omega}_k) \geq p_{\mathrm{cert}}(k/n_{\mathrm{rec}})$, uniformly over all such histories, by~\eqref{eq:recovery_lower_bound} and the monotonicity of $p_{\mathrm{cert}}$. Applying these bounds successively yields the first inequality in~\eqref{eq:success_probability}. The second follows by retaining only the factor~$k=n_{\mathrm{rec}}$, since all factors lie in $[0,1]$. \hfill$\blacksquare$
\end{pf}

The probability in Theorem~\ref{theo:algorithm_success} concerns whether the calibration finds a candidate that passes validation. It is distinct from the confidence level~$1-\beta$, which concerns the correctness of the returned bound~$\delta^*$. The lower bound~\eqref{eq:success_probability} is informative whenever the solution at $\omega=1$, which minimizes the CVaR of the constraint value, attains a non-positive CVaR, and it strengthens as this value decreases. If even this solution has a positive CVaR, the bound becomes vacuous, which reflects the control problem rather than the calibration: the latter can identify an appropriate risk weight but cannot create robustness that the problem does not admit. Since the CVaR condition is only sufficient, however, a positive CVaR at $\omega=1$ does not imply infeasibility of the chance constraint, and the procedure may nevertheless certify a candidate.

Two idealizations of the synthesis step remain. First, it solves the scenario OCP~\eqref{eq:scenario_OCP} instead of minimizing the population objective~\eqref{eq:population_objective}. Since the CVaR~\eqref{eq:cvar} is a minimum of an expectation over~$\eta$, the scenario OCP is a standard sample-average approximation in the joint variable~$(\bm{u},\eta)$, so that, under mild regularity conditions, its objective converges uniformly to the population objective as $K_{\mathrm{train}} \to \infty$~\cite{shapiro2021}. Hence, for sufficiently large training sets, the synthesized solution attains a strict CVaR margin whenever the population problem at $\omega=1$ admits one, which yields a corresponding positive lower bound on the certification probability. Second, the theorem assumes that global minimizers are obtained, which cannot be guaranteed for the non-convex problems considered here. Since Lemma~\ref{lem:margin} applies to any input sequence, a local minimizer with CVaR margin~$\kappa$ still certifies with probability at least $p_{\mathrm{cert}}$ evaluated at that margin; only the ordering of Lemma~\ref{lem:cvar_monotone} is lost. As established at the beginning of this section, neither idealization affects the validity of the certificate.

\section{Receding-Horizon Implementation}
\label{sec:receding_horizon}

The synthesis--validation procedure of Section~\ref{subsec:adaptive_calibration} computes a certified input sequence for the entire horizon from the data available at time~$t_0$. The measurements collected during execution, however, provide additional information about the system that a fixed input sequence cannot exploit, so that later inputs are chosen more conservatively than necessary. We therefore embed the procedure in a receding-horizon scheme and discuss how the resulting repeated inference and optimization can be implemented efficiently.

Let $t_i = t_0 + iL$, $i \in \mathbb{N}^0$, denote the start of the $i$th control cycle, where $L \leq H$ is the number of inputs applied per cycle, and let $\mathbb{D}_i = (\bm{u}_{0:t_i-1}, \bm{y}_{0:t_i-1})$ denote the data available at that time. In each cycle, the synthesis--validation procedure is executed with $\mathbb{D}_i$ in place of $\mathbb{D}$, the first $L$ inputs of the certified sequence are applied, and the next cycle starts from the extended dataset~$\mathbb{D}_{i+1}$.

\begin{rem}[Closed-loop guarantees]
	The guarantees of Theorem~\ref{theo:robustness_guarantees} apply to the planned trajectory of each control cycle and are re-established after every replanning step. They do not, however, constitute a joint guarantee over the complete closed-loop trajectory. This interpretation is standard in stochastic model predictive control, where probabilistic guarantees are inherently tied to the current planning problem~\cite{mesbah2016}. If a joint guarantee over a fixed number of replanning steps is required, the risk level~$\alpha$ and the confidence parameter~$\beta$ can be allocated across these steps.
\end{rem}

The computational cost of a control cycle is dominated by posterior sampling and by the synthesis of candidates, which requires solving the scenario OCP~\eqref{eq:scenario_OCP} in each attempt. Since each cycle adds only $L$ new measurements to the accumulated dataset, the posterior, and with it the scenario OCP, typically changes only gradually between cycles, particularly once the dataset is large. Both computations are therefore warm-started from the previous cycle, and within a cycle, results of previous attempts are reused.

\paragraph*{Posterior sampling}
Across cycles, samples from the previous posterior provide useful initializations for sampling from the updated posterior. This favors running one chain per retained sample over thinning a single long chain (cf.~Section~\ref{sec:pmcmc}), since each chain can be continued from its previous sample with $\mathbb{D}_{i+1}$ as target, whereas a single thinned chain would carry over only one such starting point. For PMMH, the particle filter is rerun for the current parameter to obtain an updated likelihood estimate, and the proposal and the number of particles are adapted as described in Section~\ref{sec:pmcmc}; for PG, the reference trajectory is extended over the $L$ new time steps by forward simulation. A short burn-in of $K_{\mathrm{b}}' \ll K_{\mathrm{b}}$ iterations then typically suffices, and a full burn-in is needed only in the initial cycle. Within a cycle, the posterior is unchanged, but each further attempt requires a new validation set that is independent of the previous ones (Assumption~\ref{as:independent_control_law}). Since the number of attempts typically lies below $n_{\mathrm{max}}$, these sets are generated on demand by advancing the $K_{\mathrm{val}}$ validation chains for another $K_{\mathrm{b}}'$ iterations whenever an attempt does not terminate the procedure. This amounts to thinning across attempts, with~$K_{\mathrm{b}}'$ chosen large enough that successive validation sets can be treated as independent. Apart from synchronization for adapting the PMMH proposal and number of particles, all $K_{\mathrm{train}} + K_{\mathrm{val}}$ chains evolve independently and can be run in parallel.

\begin{rem}[Relation to sequential inference]
	Sequential inference methods process new measurements incrementally and may thus appear natural for the online posterior update, but they do not provide independent posterior samples in the sense of Assumption~\ref{as:iid_posterior}. Sequential Monte Carlo methods for joint state--parameter inference, such as $\text{SMC}^2$~\cite{chopin2013}, represent the posterior by a weighted particle population whose members are mutually dependent through resampling. Forward-filtering approaches~\cite{berntorp2021} additionally never regenerate past states in light of later observations, so that path degeneracy causes their approximation of the parameter posterior to deteriorate as the dataset grows.
\end{rem}

\paragraph*{Synthesis}
A good initialization substantially reduces the time required to solve the scenario OCP~\eqref{eq:scenario_OCP}. Between control cycles, the previously returned input sequence is shifted forward by $L$ steps, padded, e.g., by repeating its final element, and used to initialize the optimization method. If state and output trajectories are also treated as decision variables, they are initialized by forward simulation of the shifted input sequence, since they depend on the newly sampled scenarios. The calibration is initialized with the risk weight of the previously returned candidate as~$\omega_1$, clipped below one to keep~$\lambda$ finite, and the most recent slope estimate as~$s_{\mathrm{init}}$, so that the first attempt typically lands close to the acceptance band~\eqref{eq:acceptance_band} and fewer attempts are needed. Within a cycle, consecutive attempts differ only in the risk weight, so the previous solution serves as the initial guess. Because the certificate holds for any candidate, these warm starts affect only the computational effort, not the per-cycle guarantees.

\section{Numerical Example: Optimal Control of a Vertical Farm}
\label{sec:example}

\begin{figure}[t]
	\centering
	\definecolor{frameGray}{RGB}{140,145,150}
	\definecolor{trayBrown}{RGB}{141,110,79}
	\definecolor{leafGreen}{RGB}{90,160,74}
	\definecolor{stemGreen}{RGB}{59,122,42}
	\definecolor{tomatoRed}{RGB}{217,72,59}
	\definecolor{ledYellow}{RGB}{245,200,60}
	\definecolor{lampGray}{RGB}{85,90,95}
	\definecolor{waterBlue}{RGB}{59,139,212}
	\definecolor{tankBlue}{RGB}{207,227,245}
	\definecolor{cylinderGray}{RGB}{176,190,197}
	\definecolor{cylinderDarkGray}{RGB}{120,144,156}
	\definecolor{climateGray}{RGB}{207,216,220}
	\tikzset{
		%leaf
		leaf/.pic={
			\fill[leafGreen] (0,0) .. controls (0.07,0.07) and (0.15,0.06) .. (0.22,0)
			.. controls (0.15,-0.045) and (0.07,-0.045) .. (0,0);
		},
		%plant
		plant/.pic={
			\draw[stemGreen, line width=0.7pt] (0,0) .. controls (0.03,0.25) and (-0.03,0.4) .. (0,0.58);
			\pic[rotate=200] at (-0.005,0.18) {leaf};
			\pic[rotate=-15]  at (0.01,0.28)  {leaf};
			\pic[rotate=195]  at (-0.01,0.40) {leaf};
			\pic[rotate=-25]  at (0.0,0.48)   {leaf};
			\pic[rotate=30,  scale=0.6] at (0,0.57) {leaf};
			\pic[rotate=150, scale=0.6] at (0,0.57) {leaf};
			\pic[rotate=90,  scale=0.4] at (0,0.57) {leaf};
			\draw[stemGreen, line width=0.4pt] (0,0.33) -- (-0.08,0.29);
			\fill[tomatoRed] (-0.09,0.25) circle (0.04);
			\draw[stemGreen, line width=0.4pt] (0,0.43) -- (0.08,0.39);
			\fill[tomatoRed] (0.09,0.35) circle (0.04);
		},
		% lamp
		lamp/.pic={
			\shade[top color=ledYellow!55, bottom color=ledYellow!5, opacity=0.8]
			(-1.15,-0.03) -- (1.15,-0.03) -- (1.35,-1.01) -- (-1.35,-1.01) -- cycle;
			\fill[lampGray, rounded corners=0.5pt] (-1.2,0) rectangle (1.2,0.12);
			\fill[ledYellow] (-1.15,-0.035) rectangle (1.15,0);
		},
		% camera
		camera/.pic={
			\fill[lampGray, rounded corners=0.4pt] (-0.11,-0.02) rectangle (0.11,-0.15);
			\fill[black!85] (-0.05,-0.15) rectangle (0.05,-0.19);
		},
		leader line/.style={draw=black!55, line width=0.35pt},
	}
	\begin{tikzpicture}[font=\scriptsize, line cap=round]
		\def\xL{2.0}
		\def\xR{5.0}
		\def\yT{3.02}
		
		% lamps with cameras
		\foreach \y in {1.31,2.74} {
			\pic at ({(\xL+\xR)/2},\y) {lamp};
			\pic at (2.95,\y) {camera};
			\pic at (4.05,\y) {camera};
		}
		
		% irrigation
		\begin{scope}
			\clip[rounded corners=1pt] (1.30,0.0) rectangle (1.72,0.75);
			\fill[tankBlue] (1.30,0.50) rectangle (1.72,0.75);
			\fill[waterBlue] (1.30,0.0) rectangle (1.72,0.50);
		\end{scope}
		\draw[waterBlue, line width=0.4pt, rounded corners=1pt]
		(1.30,0.0) rectangle (1.72,0.75);
		\draw[waterBlue, line width=0.8pt, rounded corners=1pt]
		(1.72,0.09) -- (\xL+0.14,0.09) -- (\xL+0.14,1.77) -- (\xR-0.15,1.77);
		\draw[waterBlue, line width=0.8pt] (\xL+0.14,0.34) -- (\xR-0.15,0.34);
		
		% trays
		\foreach \y in {0.16,1.59}
		\fill[trayBrown, rounded corners=0.5pt] (\xL+0.08,\y) rectangle (\xR-0.08,\y+0.14);
		
		% plants
		\foreach \y in {0.30,1.73}
		\foreach \x in {2.4,2.95,3.5,4.05,4.6}
		\pic at (\x,\y) {plant};
		
		% CO2 cylinder
		\filldraw[fill=cylinderGray, draw=cylinderDarkGray, line width=0.4pt, rounded corners=2.5pt]
		(1.33,0.90) rectangle (1.69,1.62);
		\fill[cylinderDarkGray] (1.44,1.62) rectangle (1.58,1.74);
		\draw[cylinderDarkGray, line width=0.8pt] (1.51,1.74) -- (1.51,1.95) -- (\xL,1.95);
		
		% climate unit
		\filldraw[fill=climateGray, draw=frameGray, line width=0.4pt, rounded corners=1pt]
		(2.6,\yT) rectangle (4.4,\yT+0.36);
		\foreach \y in {0.12,0.19,0.26}
		\draw[frameGray, line width=0.3pt] (2.75,\yT+\y) -- (4.25,\yT+\y);
		
		% frame
		\draw[frameGray, line width=2pt] (\xL,0) rectangle (\xR,\yT);
		
		% input labels
		\draw[leader line] (2.6,\yT+0.19) -- (2.25,\yT+0.19)
		node[anchor=east, inner sep=1pt] {Heating ($T_t$)};
		\draw[leader line] (\xL+0.31,2.54) -- (1.2,2.54)
		node[anchor=east, align=right, inner sep=1pt] {Lighting ($R_t$)};
		\draw[leader line] (1.33,1.26) -- (1.2,1.26)
		node[anchor=east, align=right, inner sep=1pt] {{CO\textsubscript{2}} ($C_t$)};
		\draw[leader line] (1.30,0.31) -- (1.2,0.31)
		node[anchor=east, align=right, inner sep=1pt] {Irrigation ($D_t$)};
		
		% crop and measurement labels
		\draw[leader line] (4.62,1.84) -- (\xR+0.25,2.10)
		node[anchor=west, align=left, inner sep=1pt]
		{Crop\\[1pt]
			state $\bm{x}_t \in \mathbb{R}^3$\\[1pt]
			parameters $\bm{\theta} \in \mathbb{R}^6$};
		\draw[leader line] (4.16,1.18) -- (\xR+0.25,0.85)
		node[anchor=west, align=left, inner sep=1pt]
		{Camera\\[1pt] $y_t = m_{\mathrm{B},t} + w_t$};
	\end{tikzpicture}
	\caption{Schematic of the vertical farm. The inputs (temperature, drought level, lighting, and CO\textsubscript{2} concentration) act on the crop, whose state~$\bm{x}_t$ evolves according to the \textit{SIMPLE} model with uncertain parameters~$\bm{\theta}$. Cameras provide noisy measurements~$y_t$ of the biomass, while the remaining states are latent (AI-assisted, Claude Opus~5.5).}
	\label{fig:vertical_farm}
\end{figure}
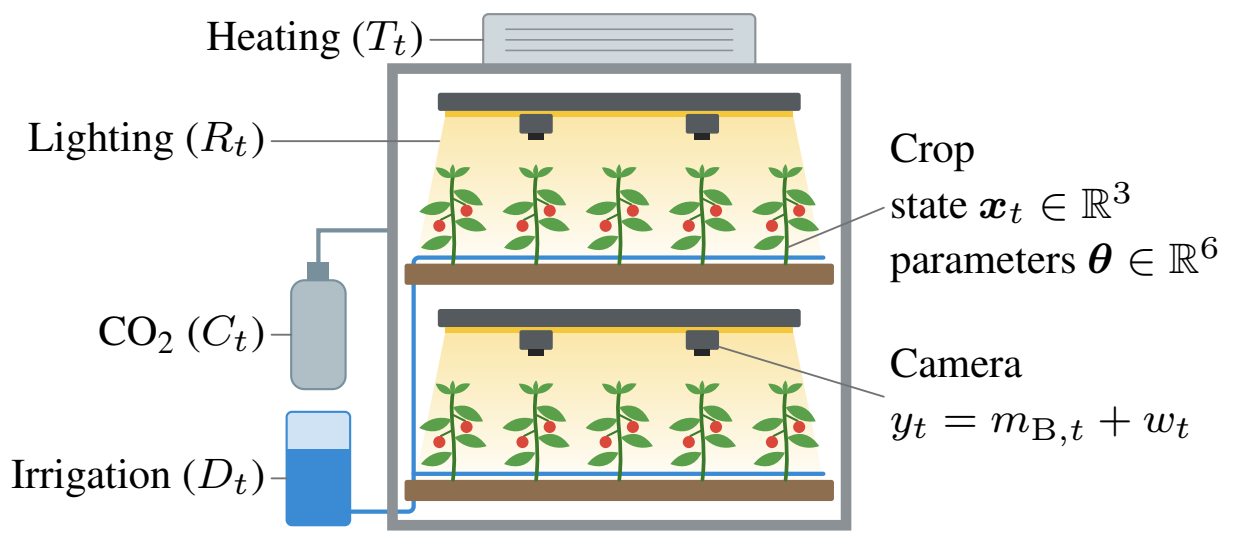

In this section, we demonstrate the proposed framework in simulation on the optimal control of a vertical farm, where crops are grown in stacked layers inside closed buildings and temperature, irrigation, lighting, and CO\textsubscript{2} concentration can be regulated independently of the weather (see Figure~\ref{fig:vertical_farm}). Since maintaining this controlled environment is costly, adapting it to the changing needs of the crop through optimal control is particularly promising~\cite{daniels2023}. We consider a problem in which these inputs are set on a daily basis to steer crop growth over a fixed cultivation period, with the goal of achieving a target yield at minimum resource cost. Meeting this target is economically important, as production may be tied to delivery commitments made before the cultivation cycle begins. Falling short of the target can lead to direct financial losses, whereas using more resources than necessary reduces the profit margin. This problem is a well-suited test case, since crop growth is described by nonlinear parametric state-space models derived from plant physiology, but the model parameters vary across cultivars and even individual plants, while only some of the physiological states can be measured, and only with considerable noise. Reaching the target yield with a prescribed probability therefore requires input plans that rigorously account for the resulting uncertainty about future crop growth. The full model specification, parameters, settings, preliminary analyses, and code to reproduce the experiments are available at \url{https://github.com/TUM-ITR/PosteriorScenarioPlanning.jl}.

\subsection{Crop Model and Prior}
\label{subsec:model}

We consider a vertical farm cultivating tomato, whose growth is described by the \textit{SIMPLE} crop model~\cite{zhao2019} in the discrete-time state-space formulation of~\cite{daniels2023} with a sampling time of one day, which matches the slow growth dynamics. The model captures the nonlinear response of crop growth to the environmental conditions and has previously been employed for optimal control in vertical farming. The state
\begin{equation}
	\bm{x}_t = \begin{bmatrix} m_{\mathrm{B},t} & T_{\mathrm{c},t} & I_{50\mathrm{B},t} \end{bmatrix}^\top
\end{equation}
contains the above-ground biomass~$m_{\mathrm{B},t}$ in \unit{\kilo\gram\per\square\meter}, the cumulative temperature~$T_{\mathrm{c},t}$, which governs the developmental stage of the crop, and the senescence variable~$I_{50\mathrm{B},t}$, which governs the onset of senescence. The input
\begin{equation}
	\bm{u}_t = \begin{bmatrix} T_t & D_t & R_t & C_t \end{bmatrix}^\top
\end{equation}
comprises the daily mean air temperature~$T_t$, the drought level~$D_t \in [0,1]$ quantifying the relative reduction in water supply, the daily radiation~$R_t$, and the CO\textsubscript{2} concentration~$C_t$. Since the inputs affect not only the current growth rate but also the timing of senescence and thus the future growth potential, planning over the entire cultivation period is required. The dynamics are augmented with additive Gaussian process noise~$\bm{v}_t \sim \mathcal{N}(\bm{0},\bm{Q})$, where $\bm{Q}$ is diagonal with standard deviations of approximately $5\,\%$ of the average daily state increments under nominal operating conditions. The states are clipped at zero to preserve nonnegativity. Only the biomass can be measured, for example using image-based methods~\cite{lu2026}, whereas $T_{\mathrm{c},t}$ and $I_{50\mathrm{B},t}$ remain latent. The measurement model is
\begin{equation}\label{eq:simple_observation}
	y_t = \begin{bmatrix} 1 & 0 & 0 \end{bmatrix} \bm{x}_t + w_t,
\end{equation}
where $w_t \sim \mathcal{N}(0,\sigma_w^2)$, with $\sigma_w$ set to approximately $5\,\%$ of the biomass after 50~days under nominal operating conditions. For simplicity of exposition, both noise covariances are assumed known, although they could be included in~$\bm{\theta}$ and estimated jointly.

Parameter values of the \textit{SIMPLE} model have been published for a range of crops and cultivars~\cite{zhao2019}, but the cultivar at hand may not be among them, and even published values may not transfer exactly due to natural biological variability and differing growing conditions. We therefore model as uncertain the cultivar-specific parameters, which differ between varieties of the same crop, and the species-specific parameters, shared by all varieties of a crop, that most strongly influence the terminal yield according to a local sensitivity analysis. All remaining parameters are fixed at their published tomato values. The resulting parameter vector~$\bm{\theta} \in \mathbb{R}^6$ governs, among other aspects, the temperature response, the conversion of radiation into biomass, and the response to drought. Since only two tomato cultivars are covered, we construct independent Gaussian priors, truncated at zero, from the sample mean and standard deviation of their values and those of five legume cultivars with similar growth characteristics, which yields a relatively broad prior. For the initial state, the cumulative temperature is zero at sowing, the initial biomass is assigned a lognormal prior with a small mean, and the prior for the initial senescence variable is constructed in the same way as for~$\bm{\theta}$.

\subsection{Control Objective and Constraints}
\label{subsec:objective}

The goal is to minimize the total production cost until harvest, which takes place $t_{\mathrm{h}} = 100$~days after sowing. At planning time~$t_0$, the inputs for the remaining $H=t_{\mathrm{h}}-t_0$~days before harvest are optimized, since the input on the day of harvest no longer affects the yield. The objective function comprises the costs of heating, irrigation, lighting, and CO\textsubscript{2} enrichment, which are linear in the inputs and based on representative prices, together with a small quadratic regularization around a nominal operating point~$\bar{\bm{u}}$ to improve numerical conditioning. The inputs are subject to box constraints representing the operating limits of the farm, with temperature and CO\textsubscript{2} concentration bounded below by their ambient values.

At harvest, a minimum yield of $Y_{\min}=\qty{8}{\kilo\gram\per\square\meter}$ must be achieved. Since the resources needed to reach this target are uncertain, the input plan must trade off production cost against the risk of missing it. As only a fraction~\mbox{$\mathit{HI} = 0.68$} of the above-ground biomass is harvestable, the target corresponds to a minimum terminal biomass of \qty{11.8}{\kilo\gram\per\square\meter}, and the requirement is expressed as
\begin{equation}\label{eq:yield_constraint}
	\mathbb{P}\left[ h(\bm{x}_{t_{\mathrm{h}}}) = \frac{Y_{\min}}{\mathit{HI}} - m_{\mathrm{B},t_{\mathrm{h}}} \leq 0 \right] \geq 1-\alpha,
\end{equation}
where $\alpha$ specifies the risk of a yield shortfall that the operator is willing to accept. For this case study, we set $\alpha=5\,\%$.

\subsection{Settings and Baselines}
\label{subsec:settings}

To assess the probabilistic guarantee empirically, we perform a Monte Carlo study with $200$ runs. In each run, the true parameters and initial state of the crop are drawn from the prior, and the farm is operated during the first $t_0=75$~days with inputs that vary randomly around the nominal operating point~$\bar{\bm{u}}$. These inputs and the resulting biomass measurements form the dataset~$\mathbb{D}$.

Posterior samples are generated by PMMH with a bootstrap particle filter, and the proposal and the number of particles are adapted as described in Section~\ref{sec:pmcmc}. Each of the $K_{\mathrm{train}}=400$ training and $K_{\mathrm{val}}=5000$ validation scenarios is obtained from a separate chain after burn-in, and fresh validation sets are generated on demand as described in Section~\ref{sec:receding_horizon}. The calibration uses $\tau=\alpha$, $\beta=10^{-3}$, $n_{\mathrm{max}}=8$, $n_{\mathrm{rec}}=3$, and $\alpha_{\mathrm{tol}}=1.5\,\%$, while $\omega_1$ and $s_{\mathrm{init}}$ are chosen based on a preliminary study, which also confirms that the power law~\eqref{eq:power_law} is a good local approximation. The scenario OCPs are implemented in JuMP~\cite{lubin2023} and solved with IPOPT~\cite{wachter2006} using the HSL linear solvers~\cite{hsl2026}. Since IPOPT requires twice continuously differentiable problem functions, all maximum and minimum operators in the crop model, including the clipping of the states at zero, are replaced by smooth approximations.

We compare the proposed framework with two baselines and a hindsight oracle, all evaluated on the same realizations of the true parameters, initial states, and noise sequences. The certainty-equivalent (CE) planner uses the same data as the proposed method but ignores the remaining uncertainty. It predicts crop growth deterministically without process noise, based on the posterior mean of the parameters, estimated from the PMMH samples, and a particle-filter estimate of the current state conditioned on these parameters. The yield requirement is then enforced on this prediction. The prior-based planner, in contrast, accounts for the uncertainty but ignores the measurements. It applies the same synthesis--validation procedure to scenarios drawn from the prior and propagated to the planning time under the applied inputs, which yields an analogous guarantee with respect to the prior predictive distribution. Finally, the hindsight oracle knows the true parameters, the current state, and the future noise realizations, and computes the cheapest input sequence that reaches the target, providing a cost reference for each run.

\subsection{Open-Loop Planning}
\label{subsec:open_loop}

\begin{figure}[t]
	\pgfplotsset{width = \columnwidth, compat = 1.18, 
		height = 0.65\columnwidth, grid = major, 
		legend cell align = left, ticklabel style = {font=\scriptsize},
		every axis label/.append style = {font=\scriptsize}
	}
	\def\file{data/seed_7_single_run.txt}
	
	\centering
	\begin{tikzpicture}
		\begin{axis}[
			grid=both,
			xmin=75, xmax=100,
			xtick distance=5,
			table/x=day,
			ylabel={$m_{\mathrm{B}}$ in \unit{\kilo\gram\per\square\meter}},
			xlabel={$t$ in \unit{\day}},
			set layers=standard,
			transpose legend,
			legend columns=4,
			legend style={font=\scriptsize, at={(0.5,-0.2)}, anchor=north, row sep=-1pt, column sep=6pt, draw=none},
			legend image post style={xscale=0.5},
			ylabel shift=-4 pt,
			xlabel shift=-4 pt]
			
			\addplot[oiBlue, very thick] table[y=proposed] {\file};
			\addlegendentry{Realized (proposed)}
			
			\addplot[name path=lower, draw=none, forget plot] table[y=lower] {\file};
			\addplot[name path=upper, draw=none, forget plot] table[y=upper] {\file};
			\addplot[fill=oiSkyBlue, fill opacity=0.3, draw=none, area legend]
			fill between[of=lower and upper];
			\addlegendentry{Pred.\ interval} 
			
			\addplot[oiBlue, densely dashed, very thick] table[y=mean_final] {\file};
			\addlegendentry{Pred.\ mean (proposed)}
			
			\addplot[oiBlue, densely dotted, very thick] table[y=mean_initial] {\file};
			\addlegendentry{Pred.\ mean (initial)}
			
			\addplot[oiVermilion, very thick] table[y=ce] {\file};
			\addlegendentry{Realized (CE)}
			
			\addplot[oiVermilion, densely dashed, very thick] table[y=ce_nominal] {\file};
			\addlegendentry{Pred.\ (CE)}
			
			\addlegendimage{empty legend}
			\addlegendentry{}
			
			\addlegendimage{empty legend}
			\addlegendentry{}
			
			\addplot[oiGreen, very thick] table[y=oracle] {\file};
			\addlegendentry{Oracle}
			
			\addplot[only marks, mark=triangle*, mark size=4pt, oiBlack,
			mark options={yshift=-4pt},
			legend image post style={mark options={yshift=0pt, xscale=2}},
			restrict x to domain=100:100] table[y=target] {\file};
			\addlegendentry{Constraint}
			
			\addlegendimage{empty legend}
			\addlegendentry{}
			
			\addlegendimage{empty legend}
			\addlegendentry{}
		\end{axis}
	\end{tikzpicture}
	
	\caption{Biomass trajectories for a representative Monte Carlo run. Solid lines show the trajectories realized on the true system, dashed lines the predictions, and the shaded area the $5$--$95\,\%$ interval of the posterior predictive distribution under the certified plan. The dotted line shows the predicted mean of the initial attempt. The triangle marks the terminal biomass required to meet the yield target. The certified plan reaches the target, whereas the certainty-equivalent (CE) plan misses it despite predicting otherwise.}
	\label{fig:single_run}
\end{figure}
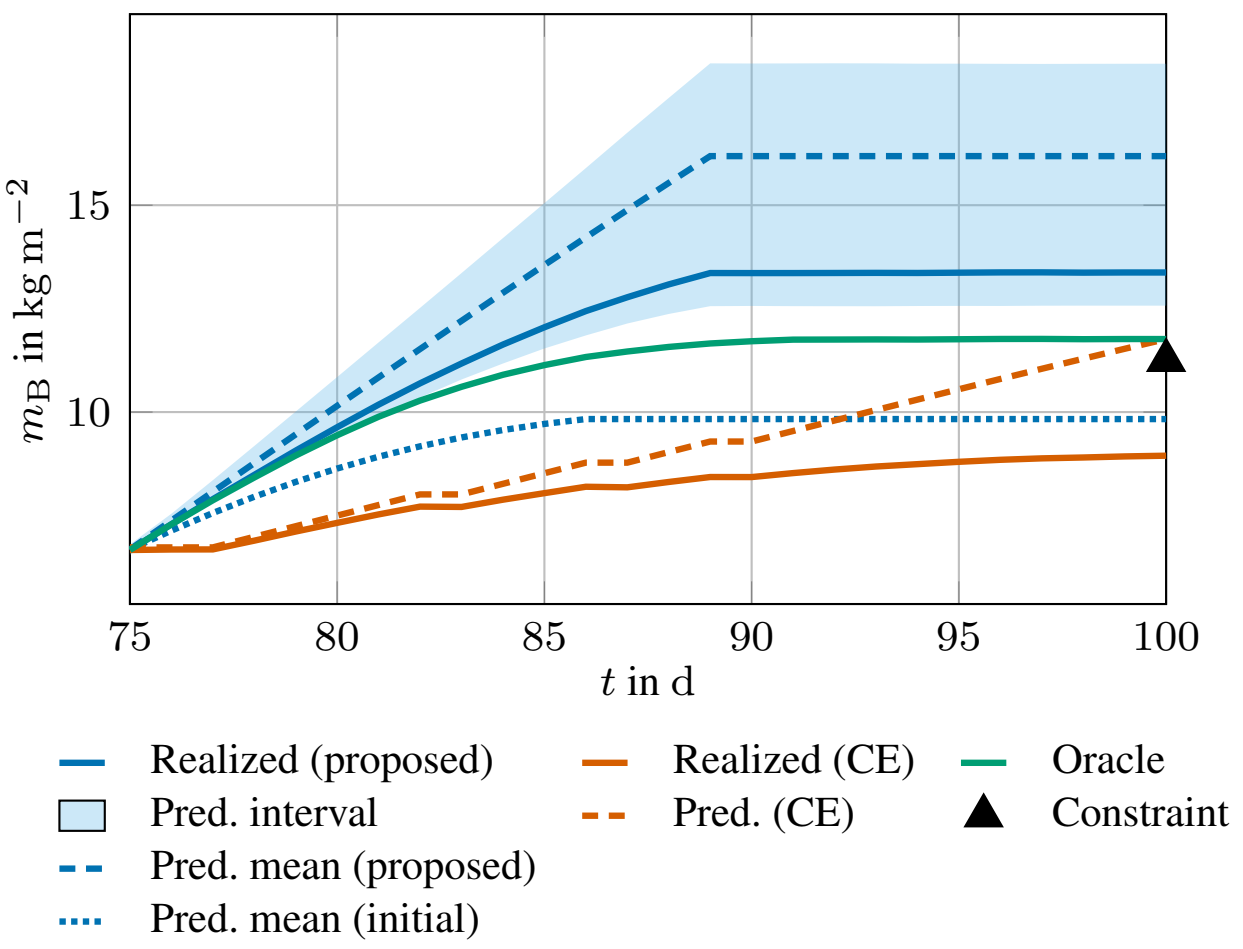

We first evaluate the framework in an open-loop setting, in which all methods plan at day~$t_0$ for the remaining \mbox{$H=25$}~days until harvest. Figure~\ref{fig:single_run} shows a representative run. The initial attempt predicts a mean biomass below the target, and the calibration raises the risk weight so that the second attempt certifies that the violation probability is at most $4.44\,\%$. The certified plan reaches the target on the true system. The CE plan predicts that the target is reached but misses it, whereas the prior-based planner fails to certify any candidate, since even its most risk-averse candidate only attains a bound of $8.26\,\%$, above $\alpha=5\,\%$.

Table~\ref{tab:open_loop_results} summarizes the results over all $200$ runs. In $8$ runs, even the oracle with perfect knowledge finds no feasible plan within the input constraints. Since the oracle problem is solved only locally, this does not strictly imply that the target is unattainable. Overall, the proposed method certifies a plan in $155$ runs, with a median certified violation bound of $4.0\,\%$. In $90\,\%$ of these runs, the bound lies within the acceptance band. None of the $155$ certified plans violates the yield constraint on the true system. Based on these independent evaluations alone, the violation probability among certified plans has a $95\,\%$ upper confidence bound of $1.9\,\%$, below the prescribed $\alpha=5\,\%$. In the remaining $45$ runs, including all $8$ without a feasible oracle plan, no candidate is certified even at maximal risk aversion, presumably because the remaining uncertainty is too large to reach the target with sufficient probability. This is the intended behavior rather than a failure: the method signals before execution that the required robustness cannot be certified, allowing the operator to react, e.g., by adjusting the target, instead of applying a plan with unknown risk. The two baselines illustrate why both learning and uncertainty quantification are required. The prior-based planner, which does not learn from the measurements, certifies no plan, as the prior uncertainty over the parameters and current state is too large. The CE planner, which neglects the remaining uncertainty, misses the target in $119$ of $191$ runs ($62.3\,\%$), far above $\alpha=5\,\%$.

The price of the guarantee is a median cost $24\,\%$ higher than that of the oracle, since the certified plan must reach the target with high probability across the possible crop realizations consistent with the data, whereas the oracle plans only for the true one. Across all $200$ runs, the synthesis--validation procedure takes on average \qty{108}{\minute} on $8$~cores of an AMD~EPYC~7402 processor, of which \qty{99}{\minute} are spent on PMCMC sampling, including the generation of fresh validation sets, and \qty{9}{\minute} on solving the scenario OCPs. Since the PMCMC chains can be run largely in parallel, the sampling time could be reduced further by allocating more cores. Even with this modest allocation, the total computation time of less than two hours is well below the one-day interval between successive control inputs.

\begin{table}[t]
	\centering
	\small
	\caption{Open-loop results over $200$ Monte Carlo runs. Plans are counted if certified (proposed and prior-based) or feasible (CE and oracle). Violations are reported among runs with a plan, and costs are medians of the per-run cost ratios to the oracle over the $155$ runs in which the proposed, CE, and oracle planners all find a plan.}
	\label{tab:open_loop_results}
	\begin{tabular}{lccc}
		\toprule
		Method & Plan found & Violations & $J/J_{\mathrm{oracle}}$ \\
		\midrule
		Proposed & $155/200$ & $0/155$ & $1.24$ \\
		Prior-based & $0/200$ & -- & -- \\
		CE & $191/200$ & $119/191$ & $1.00$ \\
		Oracle & $192/200$ & -- & $1$ \\
		\bottomrule
	\end{tabular}
\end{table}

\subsection{Shrinking-Horizon Planning}
\label{subsec:receding_horizon}

We next evaluate the receding-horizon implementation of Section~\ref{sec:receding_horizon} in the same setting. Starting at $t_0=75$, the plan is updated every $L=5$~days, resulting in $5$ control cycles until harvest. As the yield requirement refers to the fixed harvest date, the scheme is applied in its shrinking-horizon form with $H_i = t_{\mathrm{h}} - t_i$. We restrict the evaluation to the $155$ runs certified at $t_0$, as in the remaining runs the method already signals before execution that the required robustness cannot be certified. If certification fails in a later cycle, the candidate with the smallest bound is applied until the next cycle. The CE planner is likewise replanned every $L$~days and, if replanning fails, continues with the remaining portion of its last feasible plan.

\begin{figure}[t]
	\pgfplotsset{width = \columnwidth, compat = 1.18, 
		height = 0.65\columnwidth, grid = major, 
		legend cell align = left, ticklabel style = {font=\scriptsize},
		every axis label/.append style = {font=\scriptsize}
	}
	\def\file{data/seed_7_receding_horizon.txt}
	
	\centering
	\begin{tikzpicture}
		\begin{axis}[
			grid=both,
			xmin=75, xmax=100,
			xtick distance=5,
			clip=false,
			table/x=day,
			ylabel={$m_{\mathrm{B}}$ in \unit{\kilo\gram\per\square\meter}},
			xlabel={$t$ in \unit{\day}},
			set layers=standard,
			transpose legend,
			legend columns=4,
			legend style={font=\scriptsize, at={(0.5,-0.2)}, anchor=north, row sep=-1pt, column sep=12pt, draw=none},
			legend image post style={xscale=0.5},
			ylabel shift=-4 pt,
			xlabel shift=-4 pt]
			
			\addlegendimage{oiBlue, very thick}
			\addlegendentry{Shrinking horizon}
			
			\addplot[oiBlue!60, very thick] table[y=open_loop] {\file};
			\addlegendentry{Open loop}
			
			\addplot[name path=lowerA, draw=none, forget plot] table[y=lower_t0] {\file};
			\addplot[name path=upperA, draw=none, forget plot] table[y=upper_t0] {\file};
			\addplot[fill=oiSkyBlue, fill opacity=0.25, draw=none, area legend]
			fill between[of=lowerA and upperA];
			\addlegendentry{Pred.\ interval ($t=75$)}
			
			\addlegendimage{legend image code/.code={
					\draw[oiBlue!60!black, very thick] (0.3cm,-0.1cm) -- (0.3cm,0.1cm);
					\draw[oiBlue!60!black, very thick] (0.15cm,0.1cm) -- (0.45cm,0.1cm);
					\draw[oiBlue!60!black, very thick] (0.15cm,-0.1cm) -- (0.45cm,-0.1cm);}}
			\addlegendentry{Pred.\ interval ($t=95$)}
			
			\addplot[oiVermilion, very thick] table[y=ce] {\file};
			\addlegendentry{CE}
			
			\addlegendimage{empty legend}
			\addlegendentry{}
			
			\addlegendimage{empty legend}
			\addlegendentry{}
			
			\addlegendimage{empty legend}
			\addlegendentry{}
			
			\addplot[oiGreen, very thick] table[y=oracle] {\file};
			\addlegendentry{Oracle}
			
			\addplot[only marks, mark=triangle*, mark size=4pt, oiBlack,
			mark options={yshift=-4pt},
			legend image post style={mark options={yshift=0pt, xscale=2}},
			restrict x to domain=100:100] table[y=target] {\file};
			\addlegendentry{Constraint}
			
			\addlegendimage{empty legend}
			\addlegendentry{}
			
			\addlegendimage{empty legend}
			\addlegendentry{}
			
			\addplot[oiBlue, very thick, forget plot] table[y=proposed] {\file};
			
			\addplot[oiBlue!60!black, very thick, forget plot, mark=none,
			on layer=axis foreground,
			restrict expr to domain={\thisrow{day}}{100:100},
			error bars/.cd, y dir=both, y explicit,
			error bar style={very thick, oiBlue!60!black},
			error mark options={rotate=90, mark size=4pt, very thick, oiBlue!60!black}]
			table[
			x=day,
			y expr=(\thisrow{upper_last}+\thisrow{lower_last})/2,
			y error expr=(\thisrow{upper_last}-\thisrow{lower_last})/2
			] {\file};
		\end{axis}
	\end{tikzpicture}
	
	\caption{Biomass trajectories for the run of Figure~\ref{fig:single_run} under shrinking-horizon planning. Solid lines show the trajectories realized on the true system. The shaded area shows the $5$--$95\,\%$ interval of the posterior predictive biomass under the certified plan at $t=75$, while the error bar shows the corresponding interval of the terminal biomass under the certified plan at $t=95$. The triangle marks the terminal biomass required to meet the yield target. Replanning brings the terminal biomass closer to the target than the open-loop plan, whereas the certainty-equivalent (CE) planner misses the target despite replanning.}
	\label{fig:receding_horizon}
\end{figure}
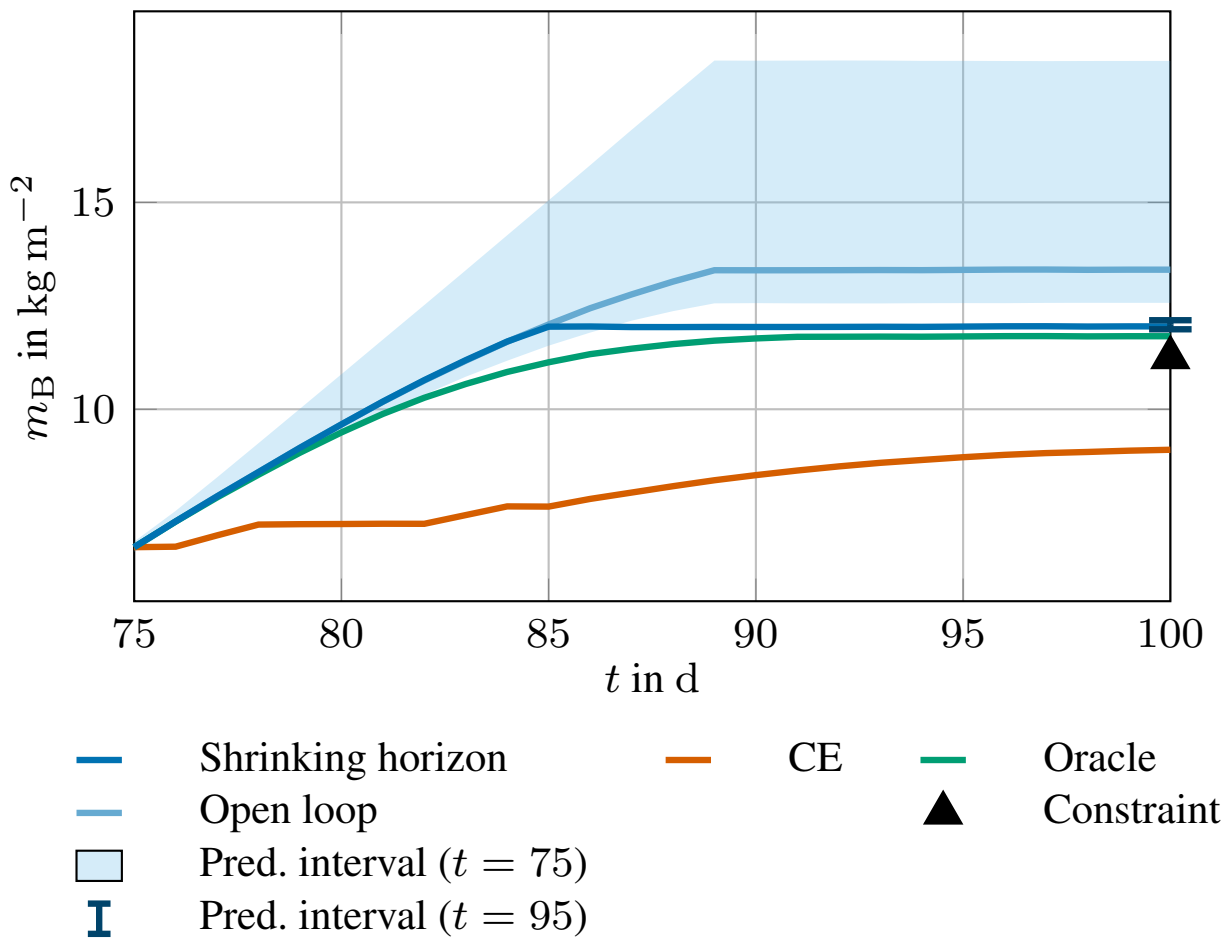

Figure~\ref{fig:receding_horizon} shows the same run as Figure~\ref{fig:single_run}. As new measurements become available, the predictive interval of the terminal biomass narrows considerably from $t_0$ to the final control cycle. The proposed planner exploits the reduced uncertainty to adapt its inputs, so that the realized yield exceeds the target by only $2\,\%$, compared with $14\,\%$ for the open-loop plan, which must hedge against the greater uncertainty at $t_0$. The CE planner, in contrast, initially plans without any margin for uncertainty. By day~$90$, the measurements reveal that the crop has fallen behind its predicted growth trajectory, and the planner fails to find an input sequence that reaches the target within the input constraints. Consequently, the target is missed despite replanning.

Table~\ref{tab:receding_horizon_results} summarizes the results. Of the $155$ runs certified at $t_0$, all subsequent control cycles are certified in $153$ runs, and none of the $155$ runs violates the yield constraint. Compared with the open-loop results of Section~\ref{subsec:open_loop}, replanning reduces the median cost ratio to the oracle from $1.24$ to $1.14$. The CE planner, in contrast, still misses the target in $113$ of $155$ runs despite replanning.

Over the $155$ evaluated runs, the first cycle, which coincides with the open-loop planning step, takes on average \qty{89}{\minute} (compared with \qty{108}{\minute} over all $200$ runs including those without a certified plan), whereas subsequent cycles take \qty{73}{\minute}, a reduction of $17\,\%$. The mean number of synthesis--validation attempts, each requiring a fresh validation set, decreases by $9\,\%$. Despite the growing dataset, the sampling time decreases by $15\,\%$, from \qty{83}{\minute} to \qty{71}{\minute}, i.e., by more than the number of attempts, which indicates that warm-starting the PMCMC chains reduces the sampling effort per cycle. The time spent solving the scenario OCPs decreases from about \qty{6}{\minute} to about \qty{3}{\minute}, reflecting the warm-started optimization but also the shorter planning horizons.

Overall, the case study shows that uncertainty quantification and learning play complementary roles in the proposed framework. Explicitly accounting for the remaining uncertainty is necessary to satisfy the constraints reliably and enables their probabilistic certification. Learning from data, both before planning and through posterior updates during operation, makes such certification attainable without excessive conservatism and reduces the cost of robustness over time. The computational effort remains well below the one-day interval between successive control inputs.

\begin{table}[t]
	\centering
	\small
	\caption{Shrinking-horizon results over the $155$ runs certified at $t_0$. A plan is counted in each cycle if certified (proposed) or feasible (CE). Violations are reported over all runs, including those in which no new plan is found in a later cycle. Costs are medians of the per-run cost ratios to the oracle over all $155$ runs.}
	\label{tab:receding_horizon_results}
	\begin{tabular}{lccc}
		\toprule
		Method & Plan in all cycles & Violations & $J/J_{\mathrm{oracle}}$ \\
		\midrule
		Proposed & $153/155$ & $0/155$ & $1.14$ \\
		CE & $143/155$ & $113/155$ & $1.00$ \\
		\bottomrule
	\end{tabular}
\end{table}

\section{Conclusion}
\label{sec:conclusion}

This article presents a Bayesian framework for the optimal control of nonlinear systems with unknown dynamics from input--output data, providing probabilistic guarantees for constraint satisfaction. The framework combines PMCMC-based posterior inference with a validation procedure that certifies constraint satisfaction and a CVaR-penalized control synthesis whose risk weight is calibrated until the desired robustness level is certified. In a receding-horizon setting, the posterior is updated as new measurements become available, allowing the controller to progressively reduce uncertainty and improve performance. A case study on the optimal control of a vertical farm confirms the guarantees empirically and illustrates how learning from data and replanning reduce the cost of robustness. Future work includes extending the framework to continuous-time models with infrequent measurements, building on MCMC methods with embedded numerical integration~\cite{lefringhausen2025}, and synthesizing output-feedback policies, whose certification is already supported by the proposed validation procedure. The synthesis of such policies is particularly challenging because control decisions must rely on noisy outputs rather than direct access to the latent state.

\section*{Declaration of Generative AI and AI-Assisted Technologies in the Manuscript Preparation Process}

During the preparation of this work the authors used Claude (Anthropic) and ChatGPT (OpenAI), in various versions between 2025 and 2026, to improve the wording and structure of the manuscript, to discuss technical ideas, to assist in programming the simulation, and to assist in creating Figure~\ref{fig:vertical_farm}. After using these tools, the authors reviewed and edited the content as needed and take full responsibility for the content of the published article.

\bibliographystyle{plain}        % Include this if you use bibtex 
\bibliography{references}        % and a bib file to produce the 
                                 % bibliography (preferred). The
                                 % correct style is generated by
                                 % Elsevier at the time of printing.

\end{document}